\documentclass[12pt]{article}
\usepackage{amsfonts}
\usepackage{amssymb, amsmath}
\usepackage{epsfig, subfig}

\begin{document}

\title{Profile Likelihood Intervals for Quantiles in \\
Extreme Value Distributions }
\author{A. Bol\'{\i}var, E. D\'{\i}az-Franc\'{e}s, J. Ortega, and E.
Vilchis. \\
Centro de Investigaci\'{o}n en Matem\'{a}ticas;\\
A.P. 402, Guanajuato, Gto. 36000;\\
M\'{e}xico}
\date{}
\maketitle

\begin{abstract}
Profile likelihood intervals of large quantiles in Extreme Value
distributions provide a good way to estimate these parameters of
interest since they take into account the asymmetry of the
likelihood surface in the case of small and moderate sample sizes;
however they are seldom used in practice. In contrast, maximum
likelihood asymptotic (mla) intervals are commonly used without
respect to sample size. It is shown here that profile likelihood
intervals actually are a good alternative for the estimation of
quantiles for sample sizes $25 \leq n\leq 100$ of block maxima,
since they presented adequate coverage frequencies in contrast to
the poor coverage frequencies of mla intervals for these sample
sizes, which also tended to underestimate the quantile and therefore
might be a dangerous statistical practice.

In addition, maximum likelihood estimation can present problems when Weibull
models are considered for moderate or small sample sizes due to
singularities of the corresponding density function when the shape parameter
is smaller than one. These estimation problems can be traced to the commonly
used continuous approximation to the likelihood function and could be
avoided by using the exact or correct likelihood function, at least for the
settings considered here. A rainfall data example is presented to exemplify
the suggested inferential procedure based on the analyses of profile
likelihoods.
\end{abstract}


\bigskip


AMS-subject classification: 62G32, 68U20.

\section{Introduction}

According to the Fisher-Tippet theorem \cite{Fisher tippet}, only three
families of distributions are the limits for the distribution of normalized
maxima of i.i.d. random variables: Weibull, Gumbel, and Fr\'{e}chet. These
three families of Extreme Value distributions (EV) are submodels of a single
family of distributions proposed independently by Von Mises \cite{vM:vM} and
Jenkinson \cite{je:je} which is now known as the Generalized Extreme Value
distribution (GEV).

Usually large quantiles \ Q$_{\alpha }$ of probability $\alpha $ of these
distributions are of interest. Different confidence intervals for these
quantiles can be obtained depending on the model used, the GEV or a specific
subfamily of models--Fr\'{e}chet, Gumbel or Weibull. Under the selected
model, the usual procedure is to obtain asymptotic maximum likelihood (aml)
confidence intervals which are symmetric about the maximum likelihood
estimate (mle) and usually do not take into account the commonly marked
asymmetry of the likelihood surface of large quantiles in the case of small
or moderate samples and thus tend to underestimate the true value of the
quantile.

Profile likelihood intervals for quantiles have not been fully explored in
statistical literature for Extreme Value Theory and neither have their
coverage properties in the cases of small and moderate samples. In this
work, the coverage frequencies and lengths of likelihood intervals for
quantiles are explored and compared to those of aml confidence intervals
through a simulation study.

In addition, the profile likelihood intervals for the shape parameter of the
GEV were also considered and shown to have good coverage frequencies. These
intervals are of special importance since they can be used as an aid for
submodel selection.

The use of the exact likelihood function, described in the following
section, is recommended for the case of small sample sizes where a Weibull
model might be reasonable, in order to avoid maximum likelihood estimation
problems due to singularities of the corresponding density function.

As an example, a data set of yearly rain maxima collected at a monitoring
station in Michoac\'{a}n, M\'{e}xico is presented to exemplify the
likelihood based estimation procedures.

\section{\textbf{Relevant Related Statistical Concepts}}

The relative and profile or maximized likelihood functions of a parameter of
interest will be presented here. In addition, the exact or correct
likelihood function is defined as well. These functions contribute to
simplify and improve the estimation of parameters of interest such as
quantiles of Extreme Value distributions. Also, expressions for the
probability densities and distribution functions of all the models involved
are here provided, as well as for their corresponding quantiles, which are
the main parameters of interest.

The densities of the three EV families for maxima are%
\begin{align}
\text{Gumbel}& \text{:}\quad \lambda (x;\mu ,\sigma )=\frac{1}{\sigma }\exp
\left\{ -\exp \left[ -\left( \frac{x-\mu }{\sigma }\right) \right] -\frac{%
x-\mu }{\sigma }\right\} I_{\left( -\infty ,\infty \right) }\left( x\right) ,
\label{ec1} \\
\text{Fr\'{e}chet}& \text{:}\quad \varphi (x;\mu ,\sigma ,\beta )=\frac{%
\beta }{\sigma }\left( \frac{x-\mu }{\sigma }\right) ^{-\beta -1}\exp \left[
-\left( \frac{x-\mu }{\sigma }\right) ^{-\beta }\right] I_{\left[ \mu
,\infty \right) }\left( x\right) ,\ \text{and}\   \label{ec2} \\
\text{Weibull}& \text{:}\quad \psi (x;\mu ,\sigma ,\beta )=\frac{\beta }{%
\sigma }\left( \frac{\mu -x}{\sigma }\right) ^{\beta -1}\exp \left[ -\left(
\frac{\mu -x}{\sigma }\right) ^{\beta }\right] I_{\left( -\infty ,\mu \right]
}\left( x\right) ,  \label{ec3}
\end{align}%
with location, scale and shape parameters $\mu \in \mathbb{R},\ \sigma >0$
and $\beta >0$, respectively. For the Weibull and Fr\'{e}chet densities, $%
\mu $ is also a threshold parameter, since it represents an upper or lower
bound, respectively, for the support of the corresponding random variable.
Note that for $\beta <1,$ the Weibull density has a singularity at $x=\mu .$

The Generalized Extreme Value distribution (GEV) density function is%
\begin{equation}
g(z;a,b,c)=\left\{
\begin{tabular}{ll}
$\frac{1}{b}\left[ 1+c\left( \frac{x-a}{b}\right) \right] ^{-1-1/c}\exp
\left\{ -\left[ 1+c\left( \frac{x-a}{b}\right) \right] ^{-\frac{1}{c}%
}\right\} I_{(-\infty ,a-\frac{b}{c})}\left( x\right) $ & $\text{if }c<0,$
\\
$\exp \left\{ -\exp \left[ -\left( \frac{x-a}{b}\right) \right] \right\}
I_{\left( -\infty ,\infty \right) }\left( x\right) ,$ & $\text{if }c=0,$ \\
$\frac{1}{b}\left[ 1+c\left( \frac{x-a}{b}\right) \right] ^{-1-1/c}\exp
\left\{ -\left[ 1+c\left( \frac{x-a}{b}\right) \right] ^{-\frac{1}{c}%
}\right\} I_{(a-\frac{b}{c},\infty )}\left( x\right) $ & $\text{if }c>0,$%
\end{tabular}%
\right.
\end{equation}%
where $a,b,c$ are location, scale and shape parameters$,$ respectively, $%
b>0\ $and $a,c\in \mathbb{R}.$ The GEV corresponds to the Weibull, Gumbel,
or Fr\'{e}chet distributions according to whether $c$ is negative, zero, or
positive, respectively. Note that the expression given for $c=0$ is the
limit of $g(z;a,b,c)$ when $c$ tends to zero. The parameters of the EV
models and the corresponding GEV are connected through a one to one
relationship given in Table 1.


\begin{center}
{%
\begin{tabular}{|l|c|l|l|l|}
\hline
& Parameter: & Threshold/Location & Scale & Form \\ \hline
Weibull & $c\,<0$ & \multicolumn{1}{|c|}{$\mu =a-b/c$} &
\multicolumn{1}{|c|}{$\sigma =-b/c$} & \multicolumn{1}{|c|}{$\beta =-1/c$}
\\ \hline
Gumbel & $c=0$ & \multicolumn{1}{|c|}{$\mu =a$} & \multicolumn{1}{|c|}{$%
\sigma =b$} & \multicolumn{1}{|c|}{---} \\ \hline
Fr\'{e}chet & $c>0$ & \multicolumn{1}{|c|}{$\mu =a-b/c$} &
\multicolumn{1}{|c|}{$\sigma =b/c$} & \multicolumn{1}{|c|}{$\beta =1/c$} \\
\hline
\end{tabular}%
}

\medskip
{Table 1. Parameters for the EV and GEV distributions}
\end{center}

In the case of the Weibull and Fr\'{e}chet models for maxima, the threshold
is isolated in a single parameter $\mu $ that may have a clear physical
interpretation. Inferences in terms of estimation intervals for this
parameter are simpler with an EV distribution in contrast to the
corresponding threshold for the GEV, which is a function of all three
parameters $a,b,c.$

It is important to note that there exist Weibull and Fr\'{e}chet
models that are very close and practically indistinguishable from a
Gumbel model. That is, the Gumbel distribution is a limit of Weibull
distributions with parameters related as shown in Table 1. The
Gumbel model is embedded in the Weibull family of models in this
sense, as well as in the Fr\'{e}chet family (Cheng and Iles
\cite{Cheng:Cheng}).

All these models can be parametrized in terms of a quantile of interest by
direct algebraic substitution in (\ref{ec1}), (\ref{ec2}) and (\ref{ec3})
since any quantile can be expressed as a function of the other parameters as
shown in Table \ref{tabb}. Therefore, the model can be expressed in terms of
the quantile of interest which substitutes one of the remaining parameters.
For example, the Weibull model can be reparametrized in terms of $\left(
Q_{\alpha },\sigma ,\beta \right) $ instead of $\left( \mu ,\sigma ,\beta
\right) $. 

\bigskip

\begin{center}
{%
\begin{tabular}{|l|l|}
\hline
& Quantile of probability $\alpha $ \\ \hline
Weibull & $Q_{\alpha }=\mu -\sigma \left( -\log \alpha \right) ^{1/\beta }$
\\ \hline
Gumbel & $Q_{\alpha }=\mu -\sigma \log \left( -\log \alpha \right) $ \\
\hline
Fr\'{e}chet & $Q_{\alpha }=\mu +\sigma \left( -\log \alpha \right)
^{-1/\beta }$ \\ \hline
GEV & $Q_{\alpha }=\left\{
\begin{tabular}{ll}
$a-b\log \left( -\log \alpha \right) ,$ & if $c=0,$ \\
$a-\frac{b}{c}\left[ 1-\left( -\log \alpha \right) ^{-c}\right] ,$ & if $%
c\neq 0.$%
\end{tabular}%
\right. $ \\ \hline
\end{tabular}%
} \label{tabb}

\medskip
{Table 2. Quantiles for the EV and GEV distributions.}
\end{center}

The asymptotic properties of maximum likelihood estimators are invoked in
order to obtain confidence intervals for the parameters of interest. Usually
the \textbf{continuous approximation to the likelihood} function as defined
in Kalbfleisch \cite{kalbfleisch} is the one used in most statistical
textbooks to define the likelihood function for continuous random variables,
without taking notice that it is an approximation. For an observed sample of
$n$ independent continuous random variables identically distributed, the
continuous approximation to the likelihood function is
\begin{equation}
L\left( \theta ;x_{1},...,x_{n}\right) =\prod\limits_{i=1}^{n}f\left(
x_{i};\theta \right) ,  \label{continua}
\end{equation}%
where $\theta $ is the vector of parameters, and $f$ is the density function
of the selected model.

This continuous approximation to the likelihood is only valid if the density
functions do not have singularities (see Montoya et al \cite{montoya}). For
example, for a given observed sample, the joint Weibull density has a
singularity when the threshold parameter equals the largest observation, $%
\mu =x_{\left( n\right) },$ if the shape parameter $\beta $ is smaller than
one, $\beta <1.$

However, the data are \textit{always} discrete since all measuring
instruments have finite precision. Therefore, the data can only be recorded
to a finite number of decimals. Thus the observation $X$ $=x$ can be
interpreted as $x-\hbox{\small $\frac{1}{2}$}h\leq X\leq x+%
\hbox{\small
$\frac{1}{2}$}h$, where $h$ is the precision of the measuring instrument,
and so is a fixed positive number. For independent observations $%
x=(x_{1},...,x_{n})$, the \textbf{exact or correct likelihood function }$%
L_{E}$ is defined to be proportional to the joint probability of the sample,

\begin{eqnarray}
L_{E}(\theta ;y) &\propto &\prod_{i=1}^{n}P(y_{i}-\hbox{\small $\frac{1}{2}$}%
h\leq Y_{i}\leq y_{i}+\hbox{\small $\frac{1}{2}$}h)  \notag \\
&=&\prod_{i=1}^{n}\left[ F\left( y_{i}+\hbox{\small
$\frac{1}{2}$}h;\theta \right) -F\left( y_{i}-\hbox{\small
$\frac{1}{2}$}h;\theta \right) \right] ,  \label{discrete}
\end{eqnarray}%
where $F$ is the corresponding distribution function of the continuous model
in consideration.

Allowing $h=0$ implies that the measuring instrument has infinite precision
and that the observations can be recorded to an infinite number of decimals.
Since for a continuous random variable $X,$ $P(X=x;\theta )$ $=0$ for all $x$
and $\theta $, this cannot be the basis for obtaining a likelihood function.
If in contrast, one assumes that the precision of the measuring instrument
is \ $h>0,$ then conditions are required for the density function $f\left(
y;\theta \right) $ to be used as an approximation to the likelihood function
$\left(\ref{discrete}\right),$ as required by the Mean Value Integral
Theorem of Calculus. But if the density function has a singularity at any
given value of $\theta ,$ then these conditions are violated and $%
f\left(y;\theta \right) $ cannot be used to approximate the likelihood
function at that value of $\theta $ (\cite{kalbfleisch}, Section 9.4).

As Meeker and Escobar (\cite{Meeker}, p.\ 275) mention, there is a path in
the parameter space for which the continuous approximation to the likelihood
$\left( \ref{continua}\right) $ goes to infinity, in particular for the
Weibull case, when $\beta <1$ and $\mu \rightarrow x_{\left( n\right) }.$ It
should be stressed that the likelihood approaches infinity not necessarily
because the probability of the data is large in that region of the parameter
space, but instead because of a breakdown in the density approximation to
the likelihood function. There is usually, as happened with all simulations
considered here, though not necessarily always, a local maximum for this
likelihood surface corresponding to the maximum of the exact likelihood
based on the probability of the data shown in $\left( \ref{discrete}\right)
. $

A useful standardized version of a likelihood function $L\left(
\theta ;x\right) $ that will be used here, is the relative
likelihood function that has a value of $1$ at its maximum, the mle
$\hat{\theta},$ and is defined as
\begin{equation}
R\left( \theta ;x\right) =\frac{L\left( \theta ;x\right) }{L\left( \hat{%
\theta};x\right) },  \label{relativa}
\end{equation}%
so that $0\leq R\left( \theta ;x\right) \leq 1.$ Values of $\theta $ with $%
R\left( \theta ;x\right) $ close to one are more plausible than values close
to zero. A relative likelihood is easy to plot and to interpret. Likelihood
intervals or regions of $k\%$ likelihood level are obtained by cutting
horizontally this likelihood function; that is
\begin{equation}
\left\{ \theta :R\left( \theta ;x\right) \geq k\right\} ,\ \ 0\leq k\leq 1.
\label{intervaloverosim}
\end{equation}%
For example, if $k=0.15,$ under some regularity conditions, the corresponding
likelihood interval has an asymptotic approximate 95\% confidence level, using the
Chi-square limit distribution for the likelihood ratio statistic (\cite{kalbfleisch}
Section 11.3). However this result may also hold for moderate samples, and even small
samples, if the likelihood surface is symmetric about the mle. In these cases the
interval in $\left( \ref{intervaloverosim}\right) $ is called a likelihood-confidence
interval.

If the GEV model is parametrized in terms of a quantile of interest, then
the profile or maximized likelihood function of $Q_{\alpha }$ (Kalbfleisch,
1985, Section 10.3) is defined for sample $x=\left( x_{1},...,x_{n}\right) $
as
\begin{equation*}
L_{p}\left( Q_{\alpha };x\right) =\max_{b,c|Q_{\alpha }}L\left( Q_{\alpha
},b,c;x\right) .
\end{equation*}%
The corresponding relative likelihood can be calculated as in $\left( \ref%
{relativa}\right) .$ Profile relative likelihoods and their plots are very
informative about plausible ranges for the parameter of interest, in the
light of the observed sample.

In the case of the profile likelihood of the GEV shape parameter $c$, the
relative likelihood at $c=0$ is indicative of the support given by the
sample to the Gumbel model, which corresponds to $c=0$. For example if $%
R_{p}\left( c=0\right) \geq 0.5,$ the Gumbel model has moderate or high
plausibility and should definitely be considered as a possible model; its
fit to the sample should be compared with the fit of the best member of the
family of EV models suggested by the sign and value of the mle $\hat{c}.$

Summarizing, in order to make inferences about a parameter of interest, for
example a quantile, the corresponding plot of the relative profile
likelihood should be analyzed because it is very informative. Inferences
about the parameter of interest should be presented in terms of
likelihood-confidence intervals, especially in the case of small or moderate
samples. These intervals calculated for two large quantiles, $Q_{.95},$ $%
Q_{.99},$ and for the GEV shape parameter $c$ showed through simulations,
reported in the following sections, to have adequate coverage frequencies
for moderate sample sizes $\left( n\geq 50\right) $, and even for $n=25$ in
the case of Gumbel and Fr\'{e}chet models.

\section{Simulations}

For the simulation study, the samples of maxima were chosen to come from one
of the EV distributions, (or equivalently a GEV distribution) and not from a
distribution belonging to the domain of attraction of an EV. Samples were
simulated from the GEV with parameters $a=1$, $b=1$ and
\begin{equation*}
c\in \left\{
-0.5,-0.4,-0.3,-0.2,-0.1,-0.05,0,0.05,0.1,0.2,0.3,0.4,0.5\right\} ,
\end{equation*}
for sample sizes of $n=25$ and 50. Additional values of $c,$ $\pm 0.01$ and $%
\pm 0.001$ were considered as well as the previous ones, for $n=100 $ in
order to explore the cases around $c=0.$ These cases are such that there are
models from the three subfamilies of EV that are very close to each other.

Size 50 is frequently found in samples coming from meteorological
applications, and sample size 100 was chosen to explore the effect of
increasing sample size. For each value of $c$ and sample size, 10,000
samples were generated in Matlab 7.

For each sample of maxima, the mle's of the parameters $\left( a,b,c\right) $
of the GEV distribution were calculated using the continuous approximation
to the likelihood function. This is the current procedure in Extreme Value
literature. The cases where the singularities of this density caused
numerical problems for finding the local maximum (the mle) were registered
and the exact likelihood function was used then to obtain the mle's.

For each simulated sample, the corresponding EV model was selected
automatically according as $\hat{c}<-10^{-5}$ (Weibull), $|\hat{c}|<10^{-5}$
(Gumbel) or $\hat{c}>10^{-5}$ (Fr\'{e}chet). The mle's of the corresponding
parameters were obtained by maximizing the likelihood derived from (\ref{ec1}%
), $\left( \ref{ec2}\right) $ or (\ref{ec3}), accordingly, reparametrized in
terms of the quantile of interest, which worked well in most of the cases.
Only when $\hat{c}<-1$ and $\hat{\beta}<1$, it was necessary to use the
corresponding exact Weibull likelihood function, as mentioned above. These
cases were registered, since they represent cases where the continuous
approximation to the likelihood function would not have been able to produce
an mle with these EV distributions.

Using the invariance property of the likelihood function, the mle's of quantiles
$Q_{.95}$ and $Q_{.99}$ can be obtained from the mle's of the parameters of the EV or
GEV, though they were obtained directly from the corresponding likelihood function
parametrized in terms of these quantiles. From their corresponding relative
likelihoods, 15\% likelihood intervals were obtained for $c,$ $Q_{.95},$ and
$Q_{.99}$. As mentioned above, these intervals may have an approximate 95\%
confidence level in the case of moderate sample sizes, using the Chi-square limit
distribution for the likelihood ratio statistic (\cite{kalbfleisch} Section 11.3).
For each of these intervals it was checked whether they included the true value of
the corresponding parameter in order to calculate the associated coverage frequency.
For those intervals that excluded the true value of the parameter of interest, the
number of times that the interval underestimated or overestimated was registered.
Also the lengths of the intervals that covered the true value of the parameter were
registered and compared as shown in the following section. In addition, the
asymptotic maximum likelihood (aml) confidence intervals were obtained for $Q_{.95}$
and $Q_{.99}$ and their coverage frequencies were registered.

\section{Results}

Tables 3 and 4 present the coverage frequencies for $Q_{.95}$ and
$Q_{.99}$ of 15\% relative profile likelihood intervals and their
corresponding aml intervals in the case of samples of size
$n=25,50,$ and 100. Asymptotically these 15\% likelihood intervals
should have 95\% coverage frequencies. Table 5 gives the coverage
frequencies of 15\% relative profile likelihood intervals for the
parameter $c$ of the GEV model for samples of size 100 and 50. The
last two columns of this table report for each scenario the number
of samples that selected the correct EV model according to the sign
of the mle $\hat{c}$ and the number of samples where the product of
the interval endpoints was negative. These are cases where the three
EV models are plausible, since the value of $c=0$ is included in the
interval.

\begin{figure}[tbp]
\begin{center}
\subfloat[]{\includegraphics[height=6cm]{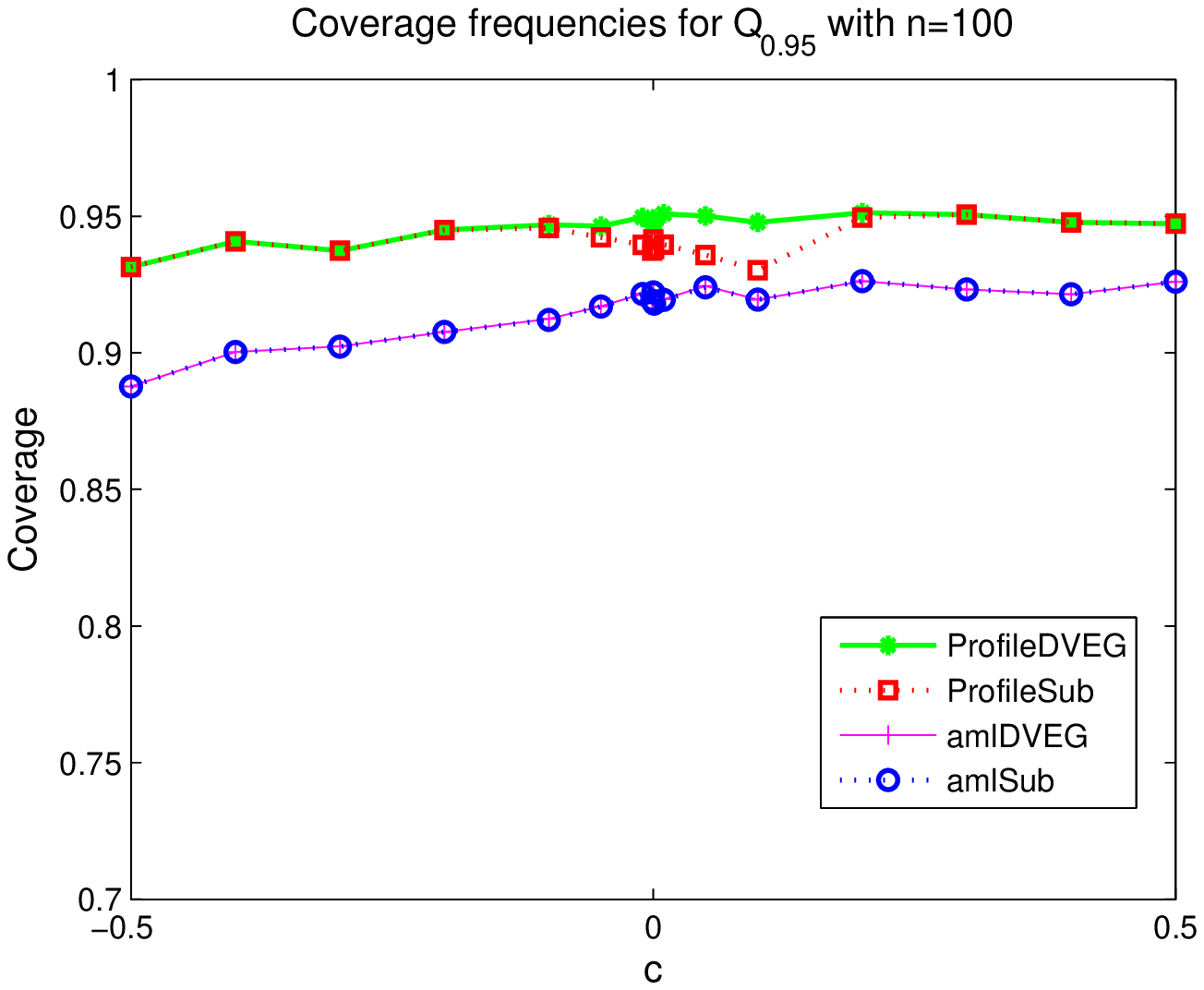}} \subfloat[]{%
\includegraphics[height=6cm]{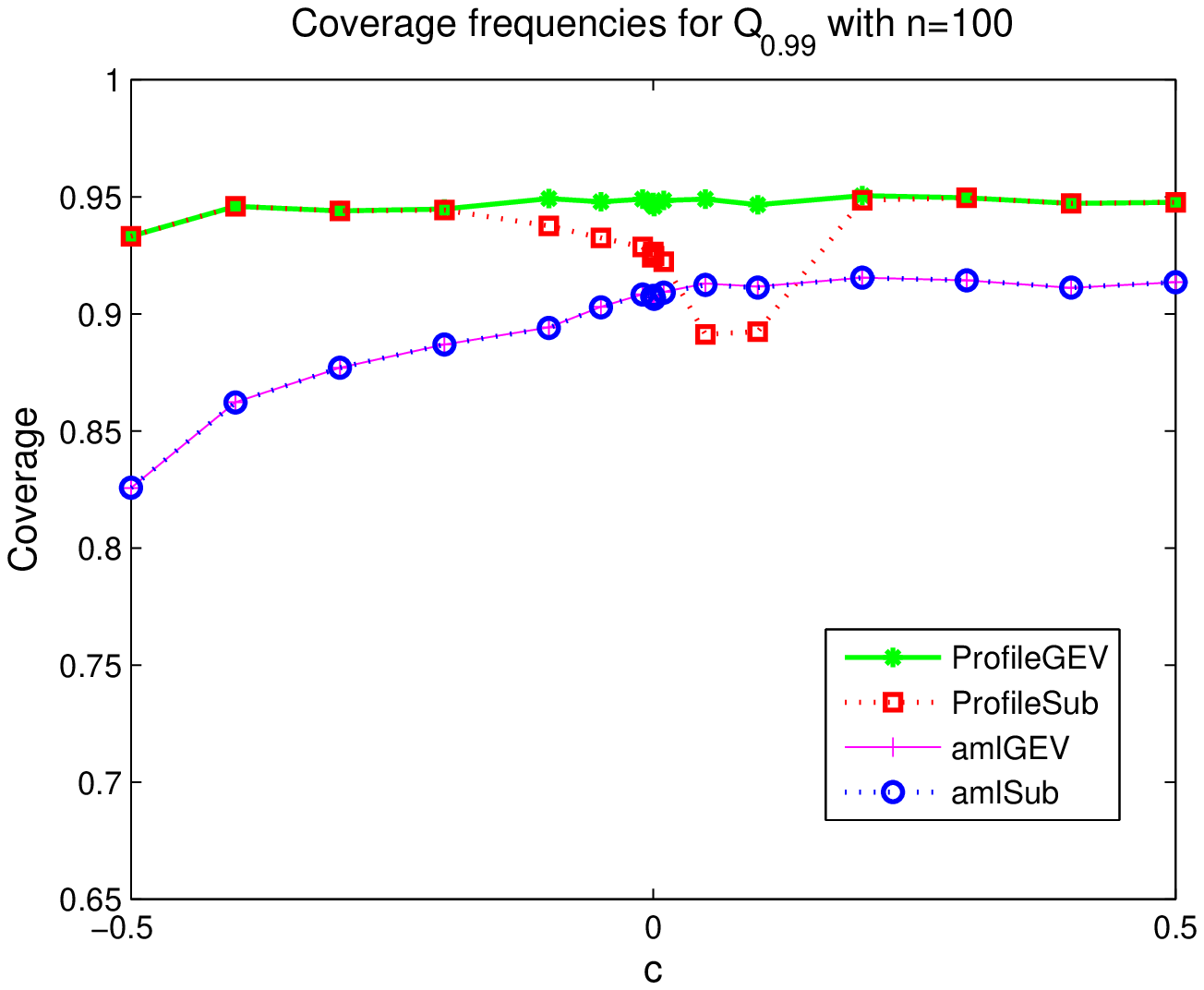}} \\[0pt]
\subfloat[]{\includegraphics[height=6cm]{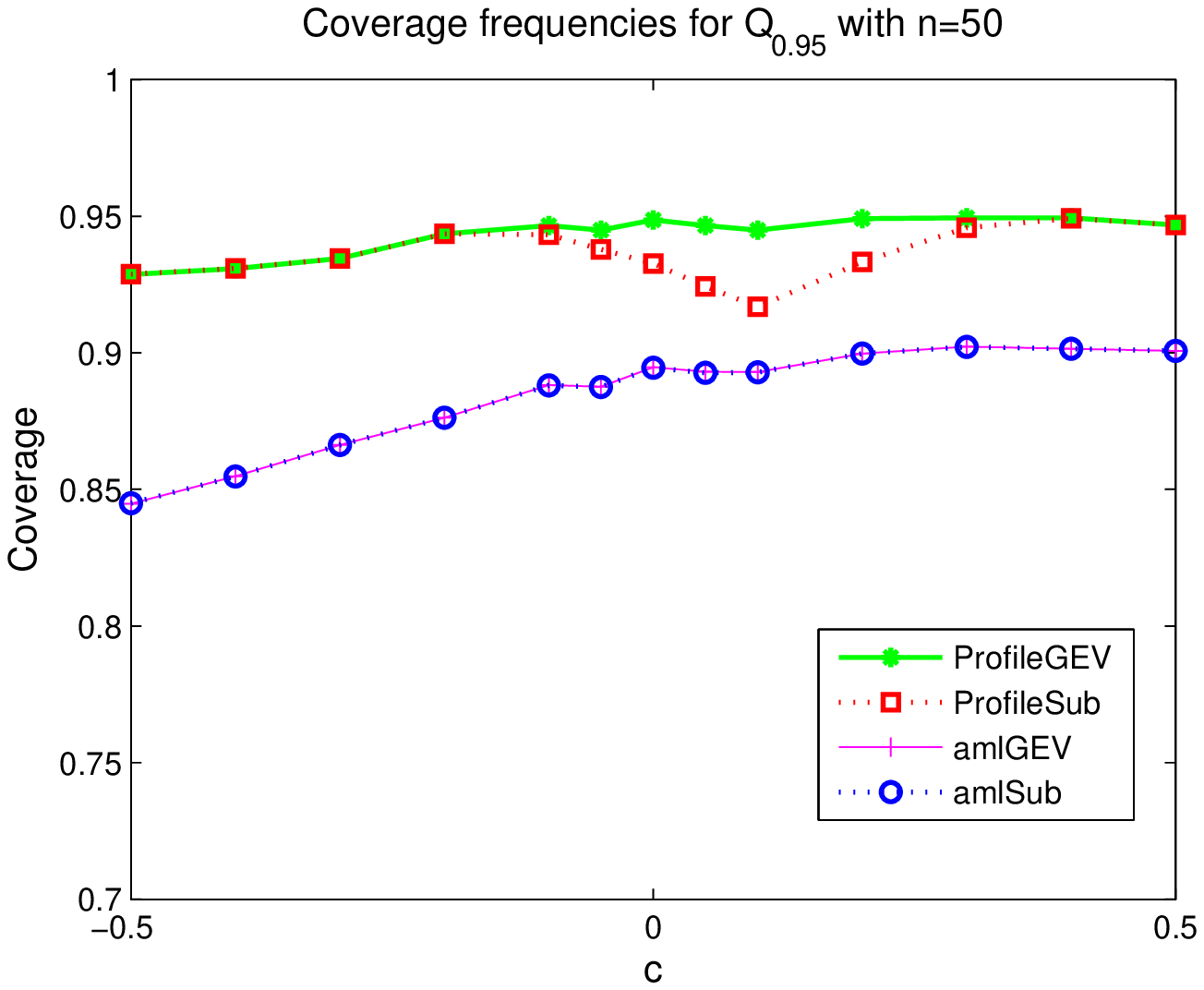}} \subfloat[]{%
\includegraphics[height=6cm]{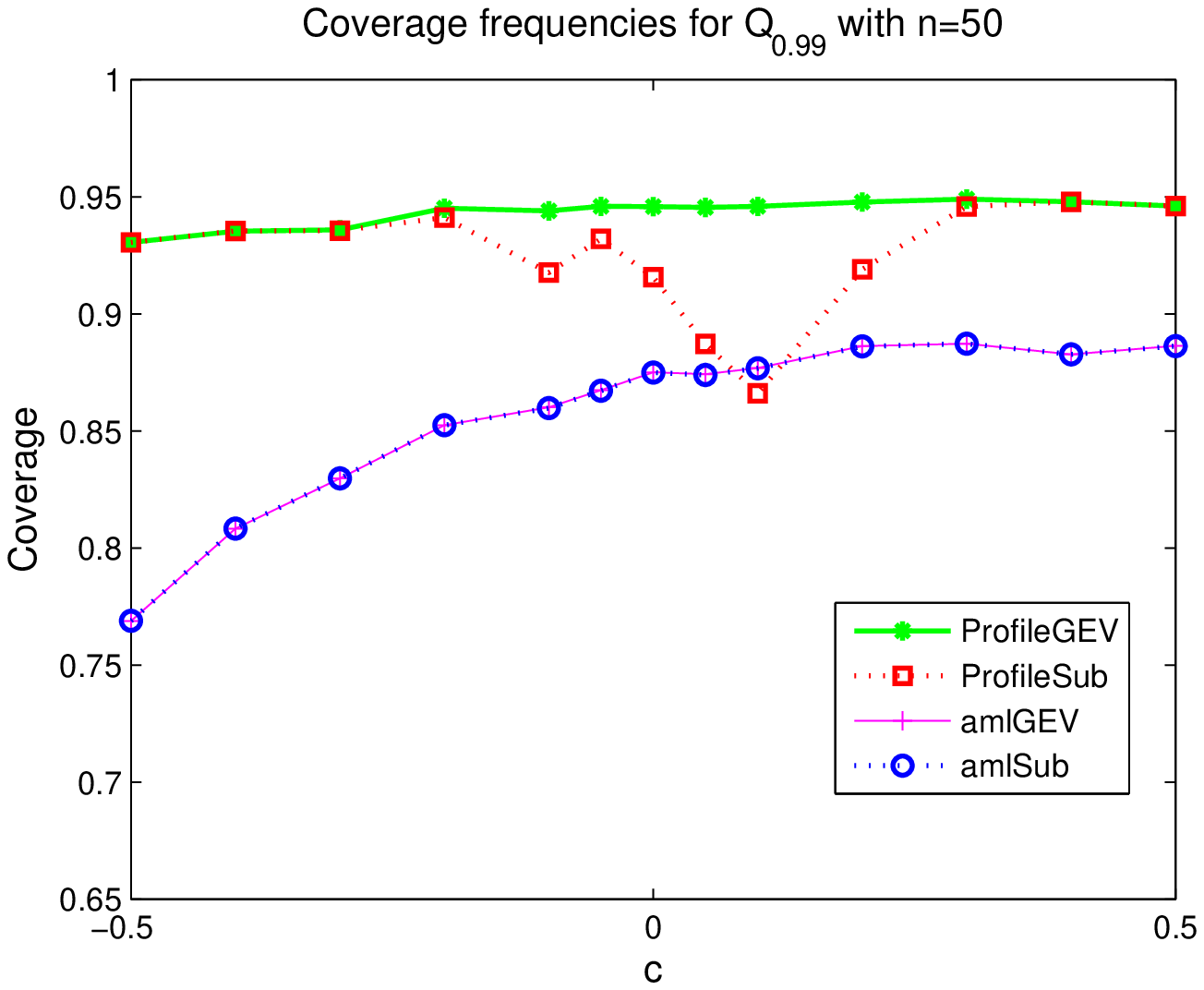}} \\[0pt]
\subfloat[]{\includegraphics[height=6cm]{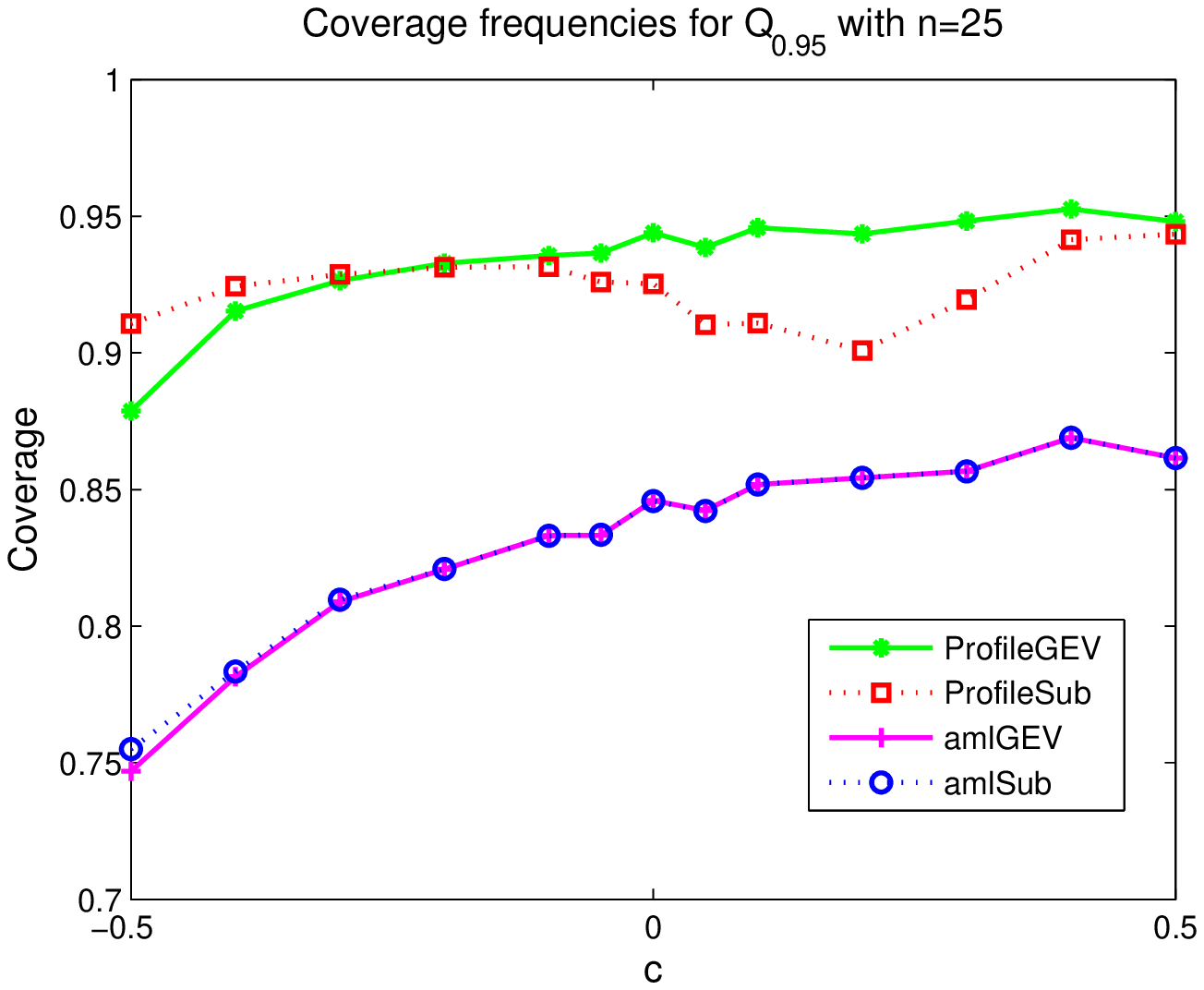}} \subfloat[]{%
\includegraphics[height=6cm]{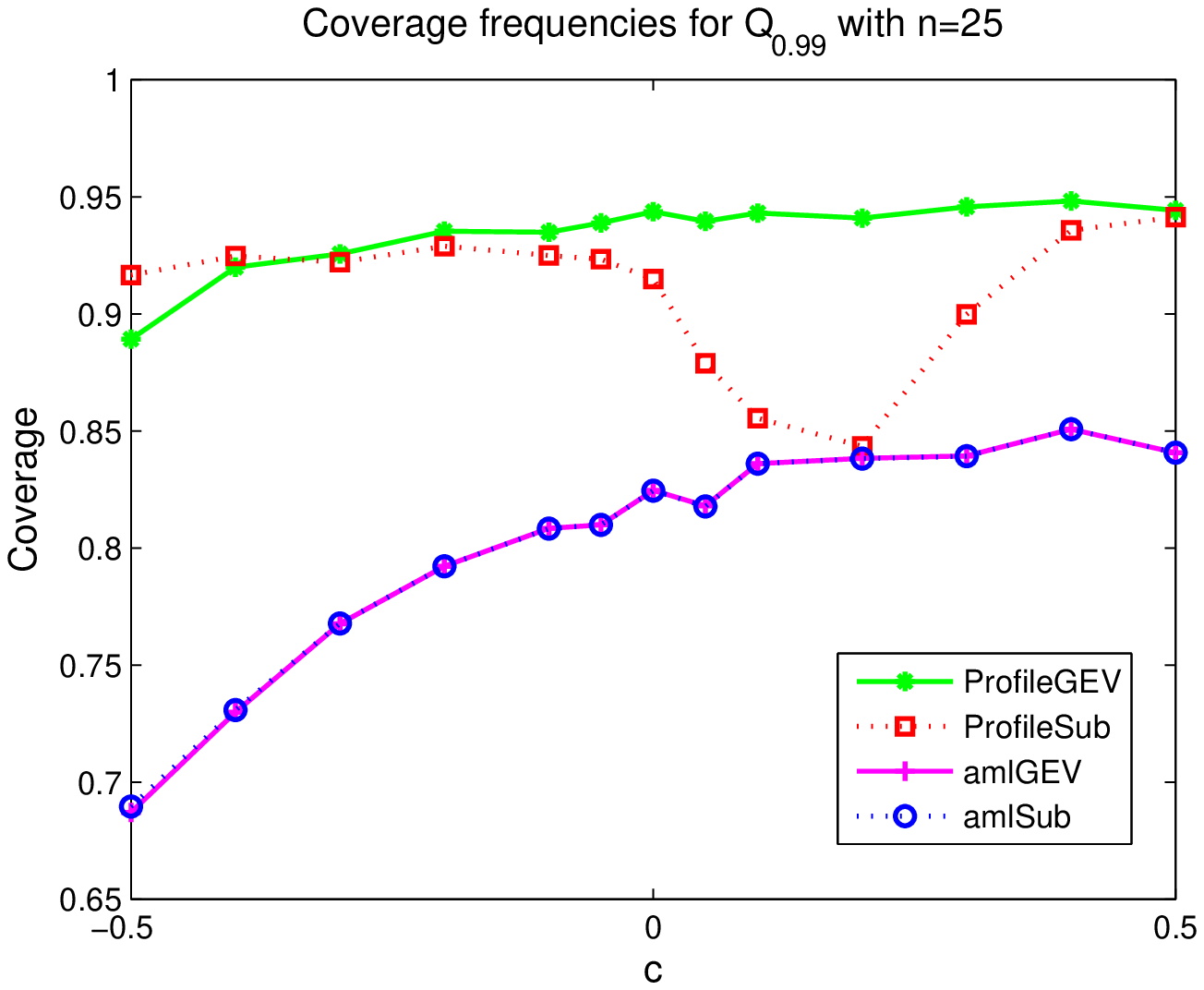}}
\end{center}
\caption{Coverage frequencies. The left column corresponds to $Q_{95}$, the
right to $Q_{99}$. The first row corresponds to a sample size of 100, the
middle row to sample size 50 and the bottom row to sample size 25.}
\label{fig1}
\end{figure}

\begin{figure}[tbp]
\begin{center}
{\includegraphics[width=12cm]{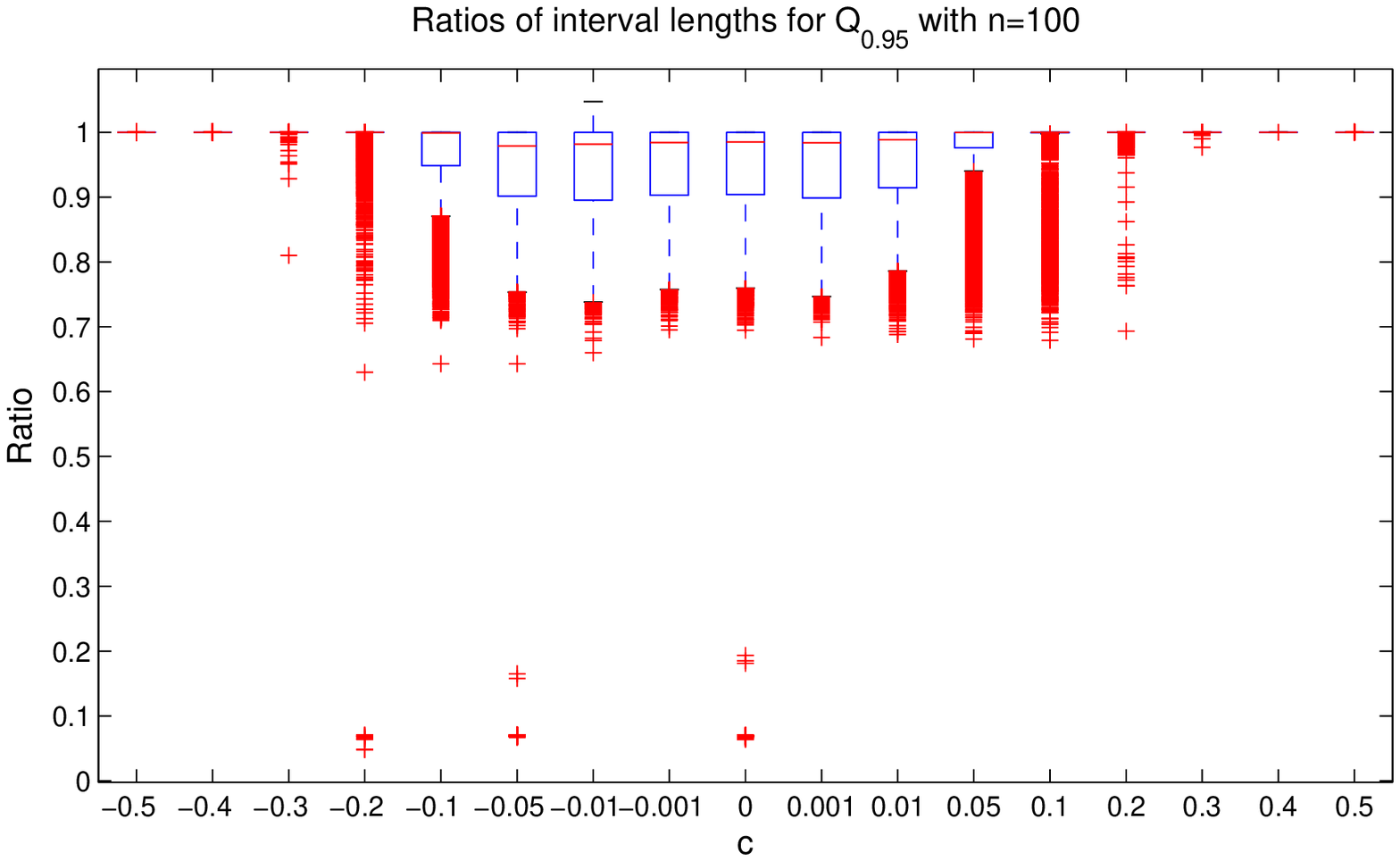}} {%
\includegraphics[width=12cm]{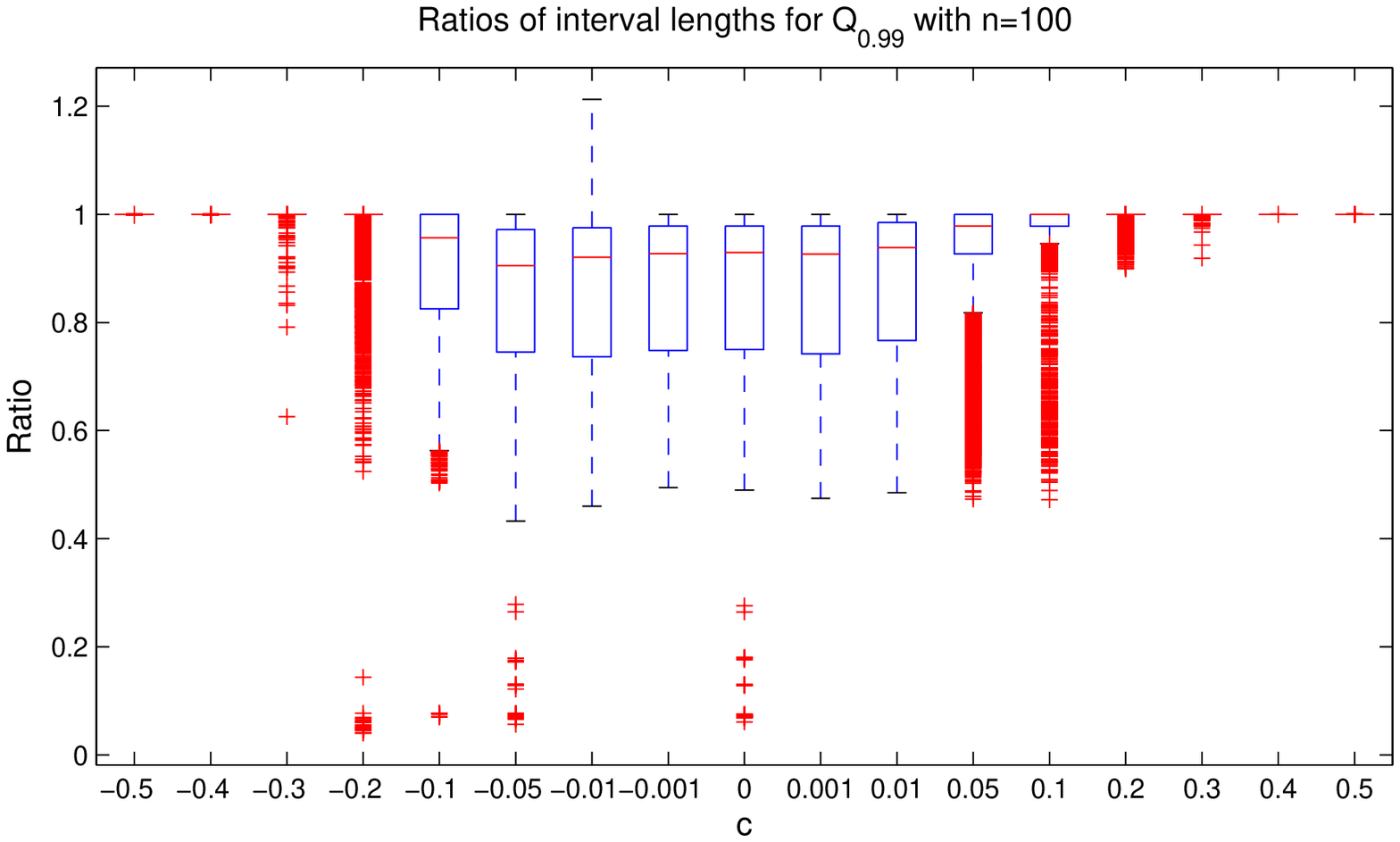}}
\end{center}
\caption{Ratio of length of likelihood-confidence intervals for $Q_{95}$
(top) and $Q_{99}$ (bottom) for the submodel over length of intervals for
the GEV, sample size 100.}
\label{fig2}
\end{figure}

\begin{figure}[tbp]
\begin{center}
\subfloat[]{\includegraphics[height=6cm]{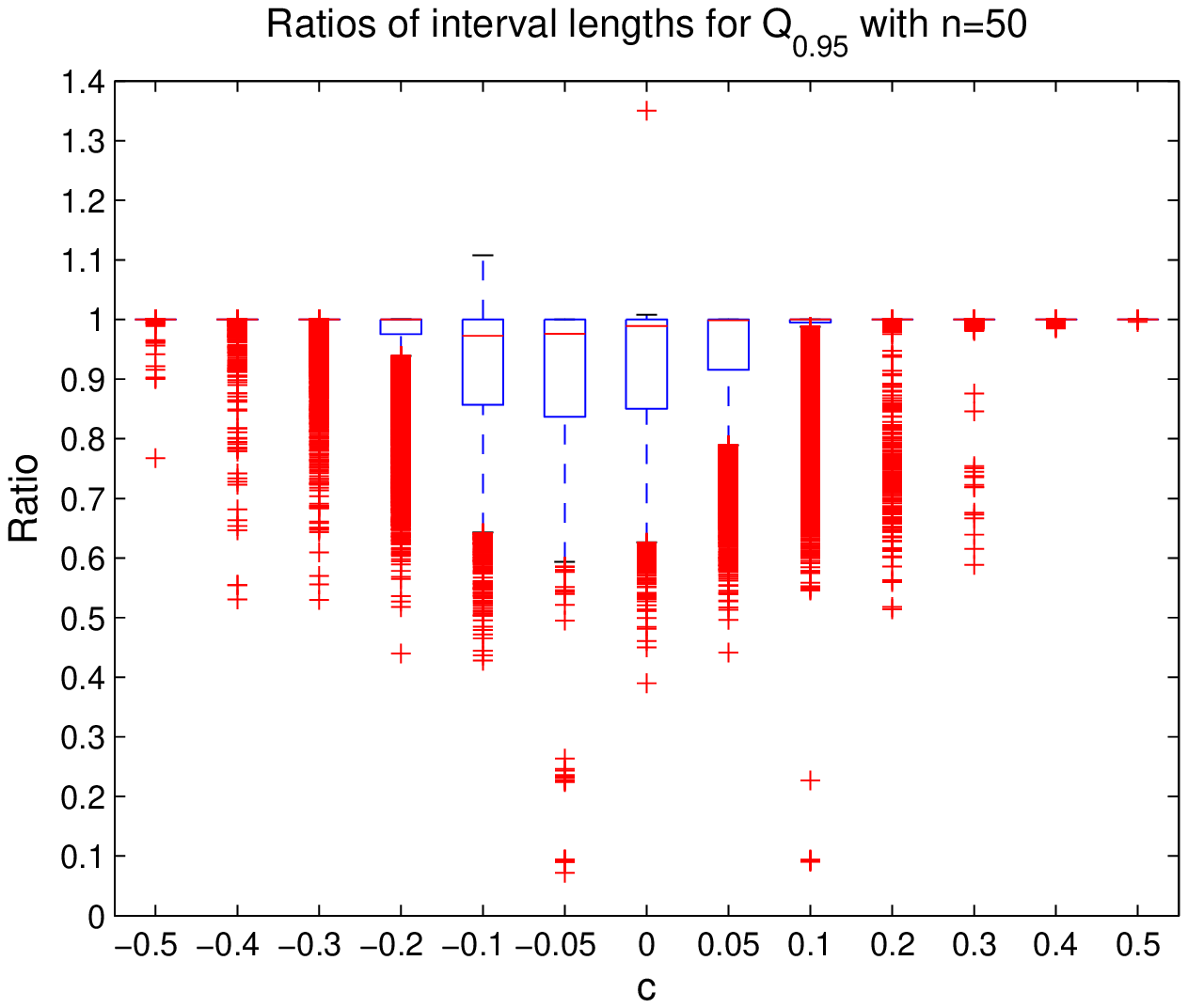}} \subfloat[]{%
\includegraphics[height=6cm]{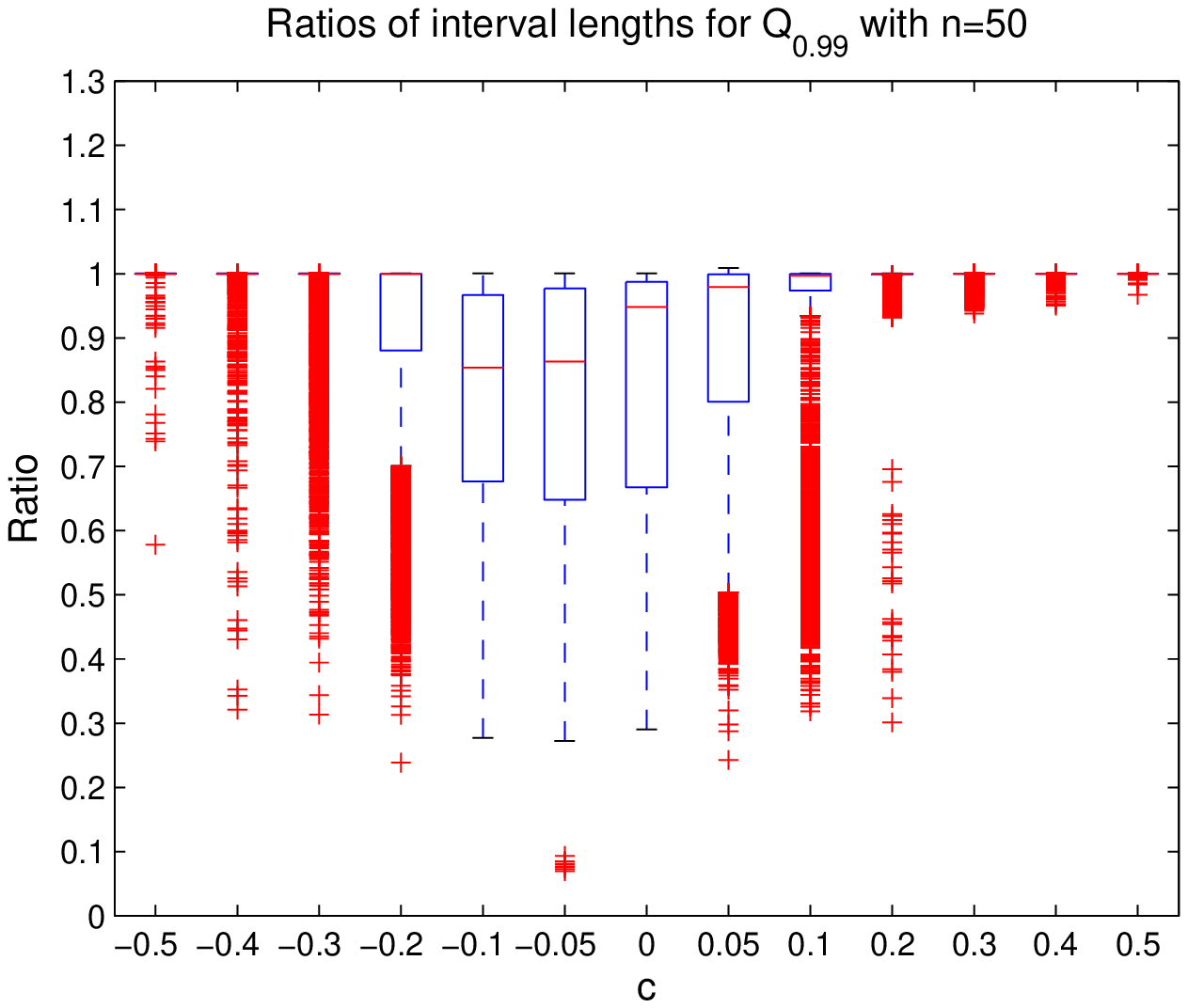}} \\[0pt]
\subfloat[]{\includegraphics[height=6cm]{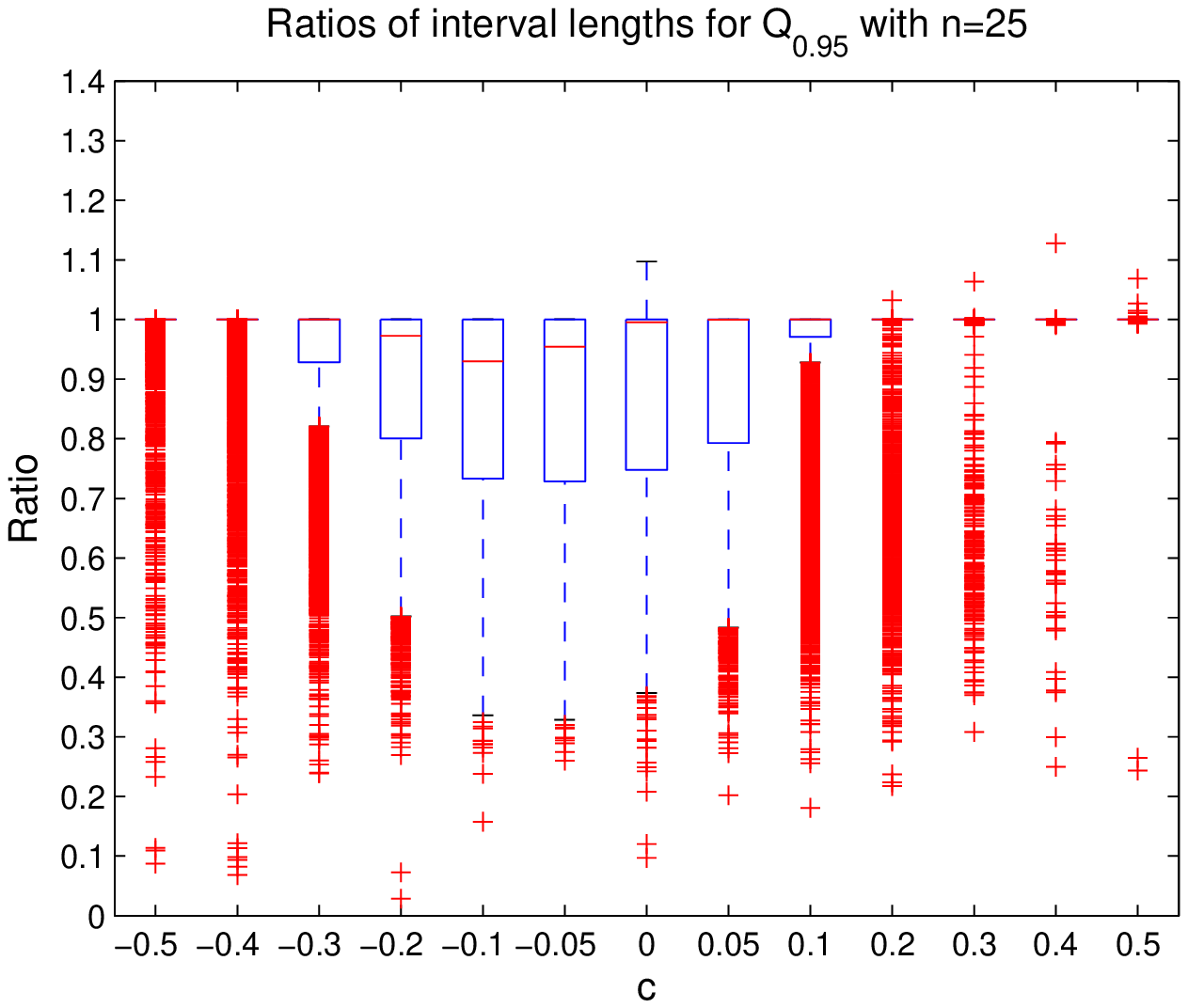}} \subfloat[]{%
\includegraphics[height=6cm]{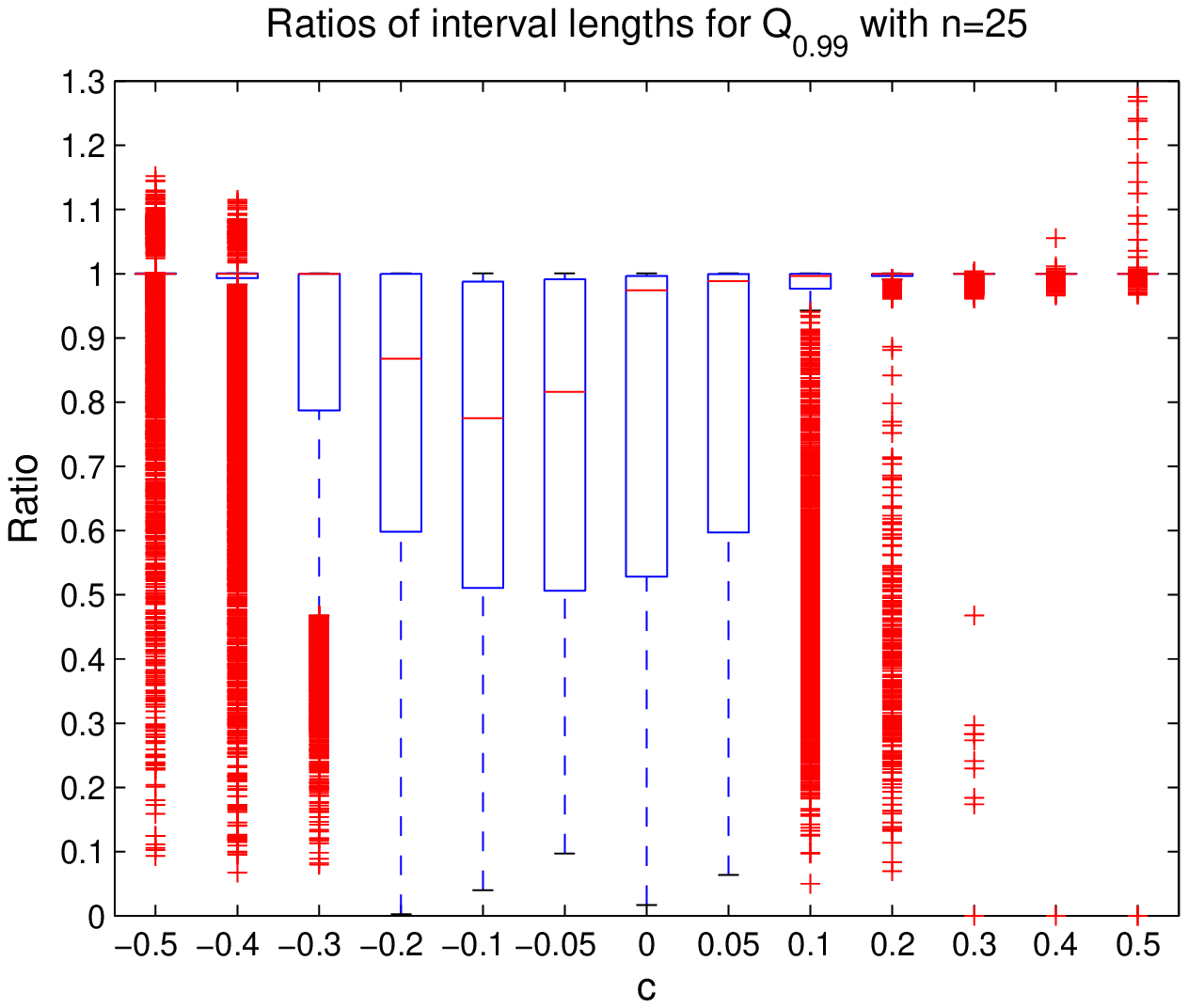}}
\end{center}
\caption{Ratio of length of likelihood-confidence intervals for $Q_{95}$
(left) and $Q_{99}$ (right) for the submodel over length of intervals for
the GEV, sample sizes 50 (top) and 25 (bottom).}
\label{fig3}
\end{figure}

\begin{figure}[tbp]
\begin{center}
{\includegraphics[width=12cm]{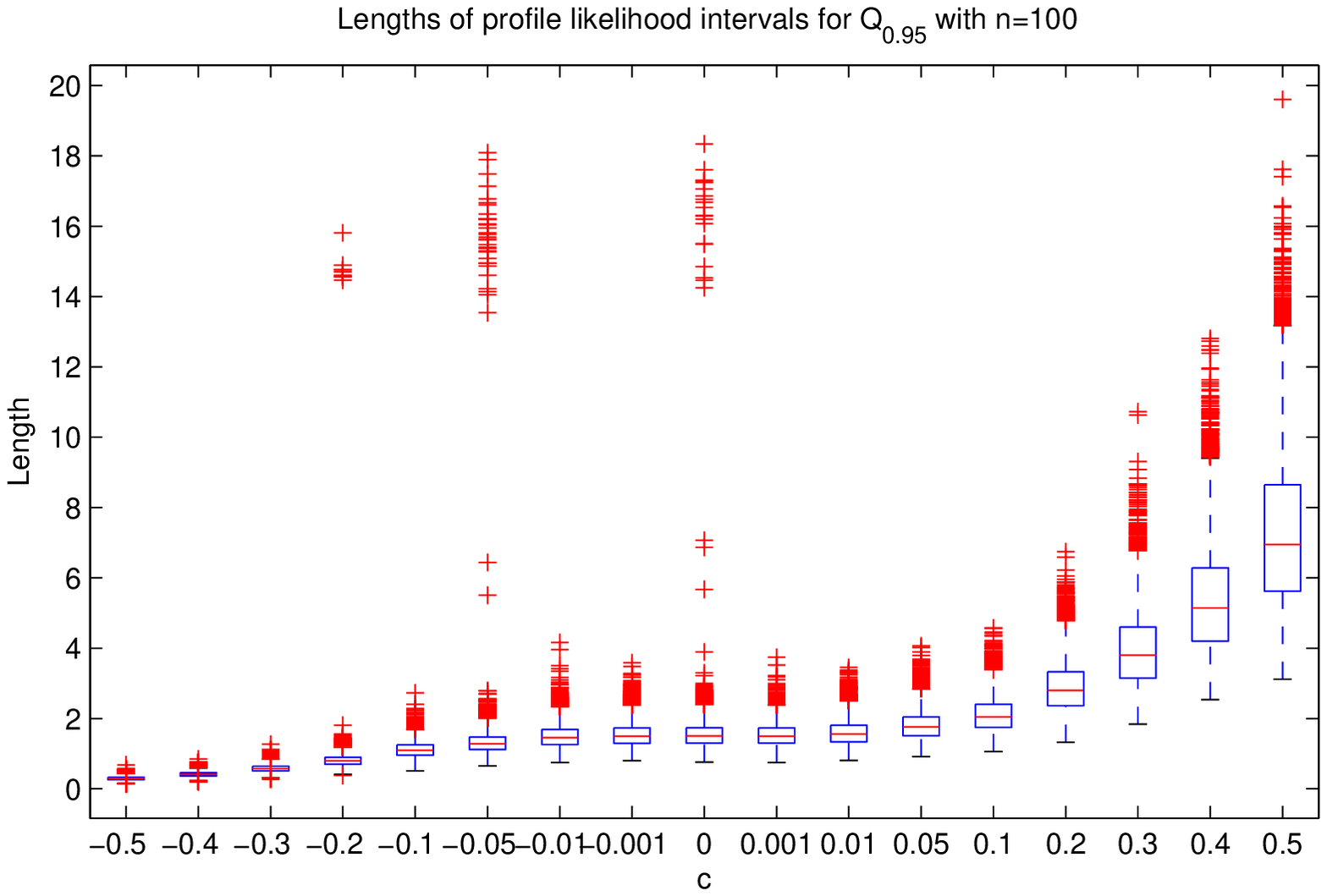}} {%
\includegraphics[width=12cm]{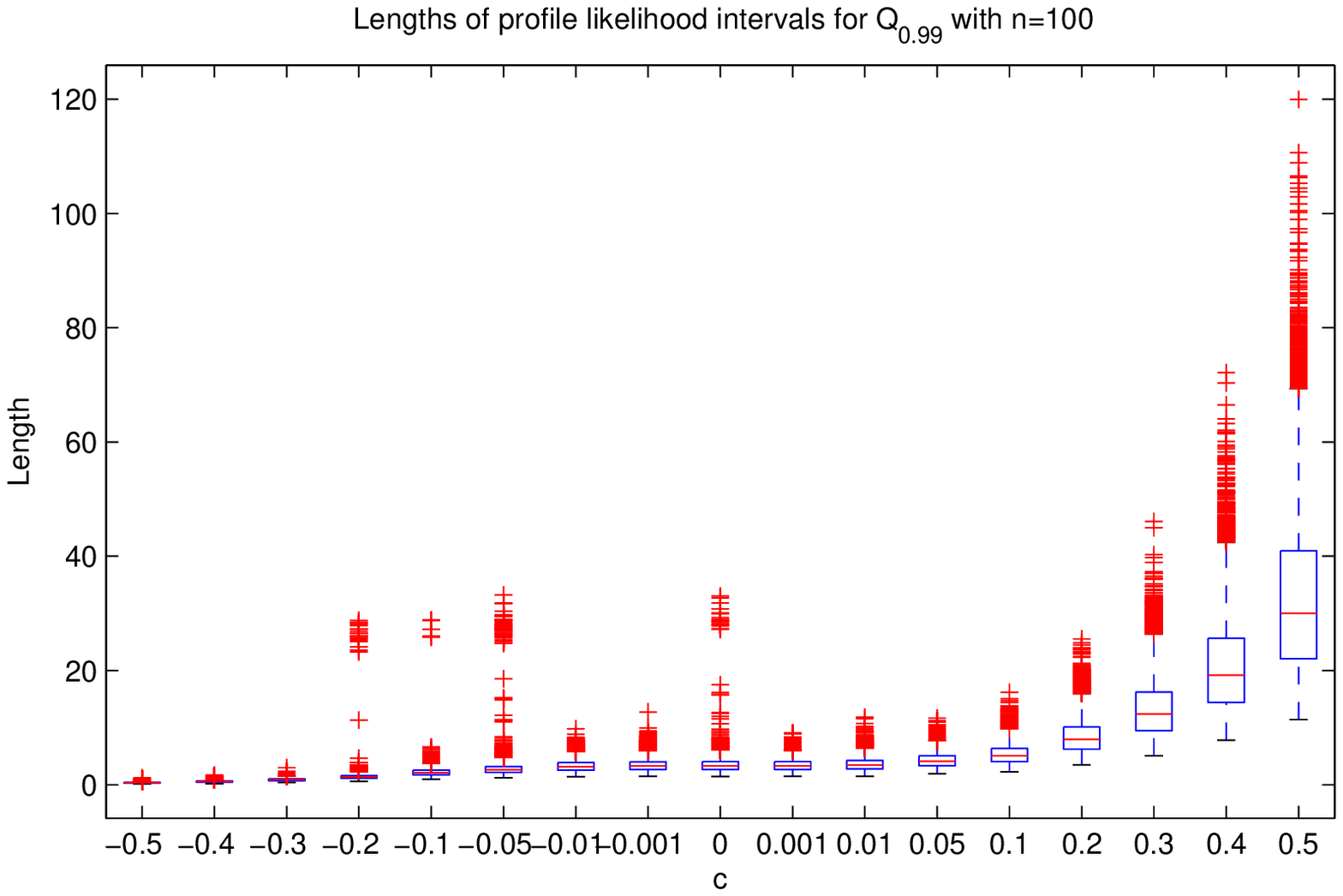}}
\end{center}
\caption{Length of profile likelihood-confidence intervals for $Q_{95}$
(top) and $Q_{99}$ (bottom) for the GEV, sample size 100.}
\label{fig4}
\end{figure}

\begin{figure}[tbp]
\begin{center}
\subfloat[]{\includegraphics[height=6cm]{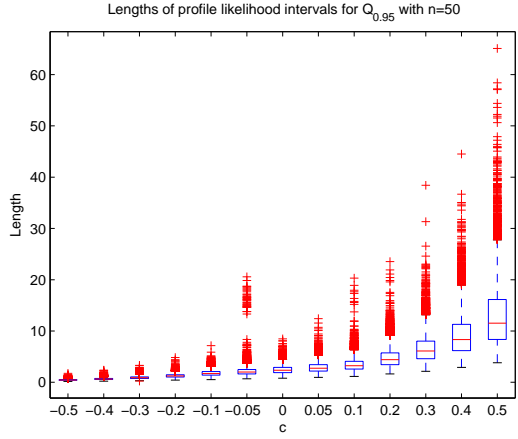}} %
\subfloat[]{\includegraphics[height=6cm]{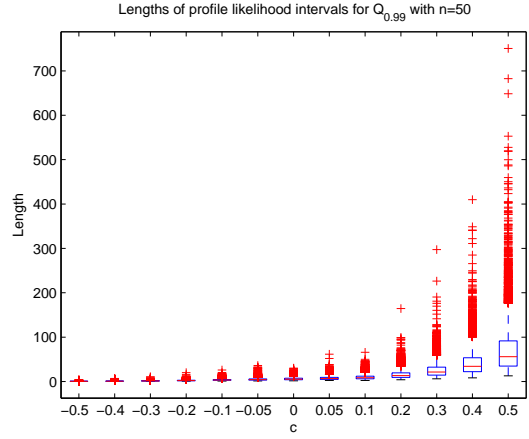}} \\[0pt]
\subfloat[]{\includegraphics[height=6cm]{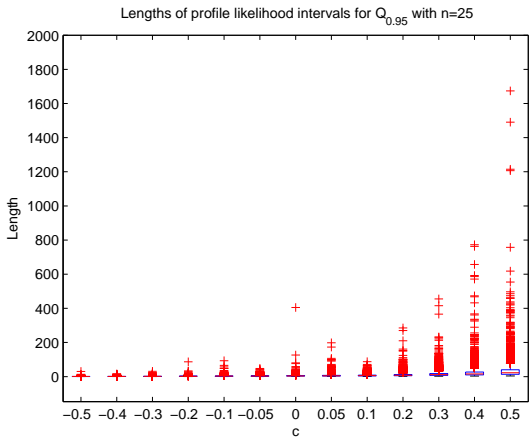}} %
\subfloat[]{\includegraphics[height=6cm]{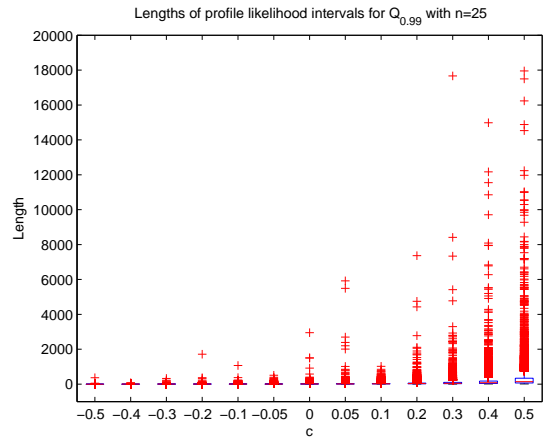}}
\end{center}
\caption{Length of profile likelihood-confidence intervals for
$Q_{95}$ (left) and $Q_{99}$ (right), sample sizes $n=50$ (top) and
$n = 25$ (bottom) for the GEV. One outlying sample was excluded from
plots (c) and (d).} \label{fig5}
\end{figure}

Figure \ref{fig1} shows the coverage frequencies of the quantiles of
interest contained in Tables 3 to 5 in a graphical way. Figures
\ref{fig2} and \ref{fig3} show the ratios of the lengths of the
relative profile likelihood intervals under the selected EV model
compared to those under the GEV model and Figures \ref{fig4} and
\ref{fig5} give the length of profile likelihood intervals for the
GEV using boxplots in which the box corresponds to the interquartile
range and the whiskers have a maximum length of 1.5 times the
interquartile range. Points beyond the end of the whiskers are
represented individually and the line inside the box is the median.
Only samples for which all intervals covered the true value of the
quantile were considered in these graphs.

Some remarks about the tables and figures are given below. Note that EV
submodels are selected automatically, based only on the sign and size of $%
\hat{c},$ so the reported coverage frequencies correspond to a `worst case'
scenario. With a real data set, additional external information from experts
would be taken into account for choosing an adequate submodel, and
consequently the statistical modeling would be more efficient.

\begin{figure}[tbp]
\begin{center}
{\includegraphics[height=6cm]{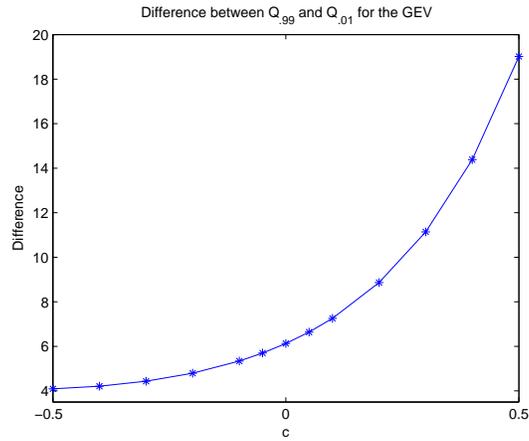}}
\end{center}
\caption{Difference between $Q_{01}$ and $Q_{99}$ for the GEV models
with $a=b=1$ and corresponding values of $c$.} \label{fig6}
\end{figure}

\begin{enumerate}
\item \textbf{Coverage frequencies of GEV profile likelihood intervals and
number of samples with estimation problems.} Coverage frequencies of
relative profile likelihood intervals for the GEV were very stable
throughout the range of values of $c$ for both quantiles. They tend to
decrease as $c$ moves towards more negative values. For $n=100$ there were
no numerical problems when calculating the mle's. For $n=50$ the number of
samples with numerical problems was insignificant. However for $n=25,$ more
samples presented problems in the case of Weibull models with values of $c$
smaller than $-0.2.$ The number of problematic cases grows as $c$ goes to $%
-0.5$ and is above $1.8\%$ for $c=-0.4$ and above $5\%$ for
$c=-0.5$. The number of samples that had numerical problems was the
same for both quantiles considered. Therefore, numerical problems
are associated to small sample sizes and Weibull models with large
negative values of $c.$

\item \textbf{Coverage frequencies of EV profile likelihood intervals.}
Coverage frequencies of relative profile likelihood function
intervals for the EV were not so stable, and in all cases there is a
region of decrease, mainly in the Fr\'{e}chet domain, where
frequencies drop, as shown in Figure \ref{fig1}. This region grows
wider as the sample size gets smaller, and the value where the
minimum
occurs shifts to the right from around 0.1 for $n=100$ to around 0.2 for $%
n=25$. The drop is always more pronounced for $Q_{.99}$ than for
$Q_{.95}$. This can be explained by the fact that for the samples
that did not cover the true value of the quantile, the mle $\hat{c}$
was negative in most cases and the whole interval lay below this
true value and therefore underestimated it (see the second columns
in Tables 3 and 4). In the Fr\'{e}chet cases, these problems were
associated to estimating a large Fr\'{e}chet quantile with a Weibull
model that has a bounded right tail.

\item \textbf{Coverage frequencies of aml intervals.} Aml intervals always
had poorer coverage frequencies than relative profile likelihood
intervals for the GEV for all the sample sizes considered here.
Coverage frequencies for aml intervals calculated for the GEV and EV
distributions are almost identical. Although coverage frequencies
for these intervals improve as the sample size grows, as predicted
by asymptotic theory, they can be very poor for $n=25$ and 50, and
still unsatisfactory even for $n=100. $ This indicates that samples
of greater size are required for these intervals to have suitable
coverage frequencies. In all cases the intervals that failed to
cover the true values tended to underestimate them.

\item \textbf{Asymmetry of proportions of intervals that exclude the true
value.} Except for one single case ($n=50,\ Q_{.99},\ c=0.5$) there
were always more relative profile likelihood intervals that
underestimated than overestimated the true value of the quantile.
This asymmetry is more pronounced for smaller sample sizes, $n=25.$
The asymmetry also increases as $c$ becomes smaller and is very
marked in the Weibull case. This may be due to the fact that the
Weibull distribution has a finite upper limit and intervals tend to
increase
in size with $c$. Therefore estimating a large quantile from a sample with $%
\hat{c}<<c$ will tend to underestimate the true value while in the case $%
\hat{c}>>c$ the interval will be larger and more likely to include the true
value. However, even if this asymmetry is not desirable, the asymmetry of
aml intervals is certainly much more marked than the one for profile
likelihood intervals.

\item \textbf{Interval lengths.} Almost always intervals obtained with the
GEV models are larger than those obtained with EV distributions as
shown in Figures \ref{fig2} and \ref{fig3}. Only samples where both
intervals included the true value of the parameter were considered.
The length of the intervals tended to be alike for large values of
$|c|$, although there is some asymmetry in this, with Fr\'{e}chet
intervals being closer in length than the corresponding Weibull
cases. Also, the ratio of lengths is closer to one for $Q_{.95}$
than for $Q_{.99}$. For both quantiles the largest difference occurs at $%
c=-0.05$ for $n=100$ and $50$, and at $c=-0.1$ for $n=25$. In
Figures \ref{fig2} and \ref{fig3}, the region where the
interquartile boxes are visible (i.e. where the length differences
are more important) coincides roughly with the region where there is
a drop in the coverage frequencies for the EV distributions. This
shows that there is a trade off between coverage and precision in
the choice of a model: There is the possibility of gaining precision
in the estimation but a the risk of reducing the confidence level of
the interval. It is important to note that for the same quantile and
sample size, the lengths of confidence intervals grow with $c$, as
shown by Figures \ref{fig4} and \ref{fig5}. This is to be expected
since Weibull distributions are bounded above while Gumbel and
Fr\'{e}chet are not. Figure \ref{fig6} shows the length between the
true values of $Q_{.01}$ and $Q_{.99}$ of the corresponding
distribution, as the parameter $c$ increases.

\item \textbf{Effect of sample size on interval length.} As one would
expect, the length of the intervals decreases as the sample size increases,
but not uniformly. Halving the sample size from $n=50$ to 25 increases
interval length by a factor between 1.84 to 2.65, depending on the value of $%
c,$ and by a factor of 1.56 to 1.78 when decreasing from $n=100$ to
50. Also, for a fixed sample size the length of intervals for
$Q_{.99}$ is always larger than those of $Q_{.95}$, as shown in
Figure \ref{fig5}.

\item \textbf{Coverage frequencies of GEV shape parameter }$c.$ The coverage
frequencies of the profile likelihood intervals of this parameter, shown in
Table 5, are stable throughout the range of values of $c$, with a slight
decrease for the more negative values of $c$. The proportion of intervals
that underestimate is much larger than those that overestimate the true
value of $c,$ especially in the Weibull cases. This asymmetry diminishes as $%
c$ takes larger positive values.

\item \textbf{Asymmetry in the correct automatic selection of a model.} The
number of simulated samples where the estimator $\hat{c}$ has the same sign
as the true value of $c,$ as the column \textquotedblleft
correct\textquotedblright\ shows in Table 5, depends on the value of $c$.
Although the difference is not pronounced, it is always more likely for the
same value of $|c|$ that the signs coincide in a Weibull case than in the
corresponding Fr\'{e}chet case. On the other hand, it is more likely that
intervals in the Fr\'{e}chet case cover the origin, and therefore make
plausible a Gumbel model, as the \textquotedblleft
negative\textquotedblright\ column shows in Table 5.
\end{enumerate}

\section{Rain Data Example}

\begin{figure}[tbp]
\begin{center}
\subfloat[]{\includegraphics[height=6cm]{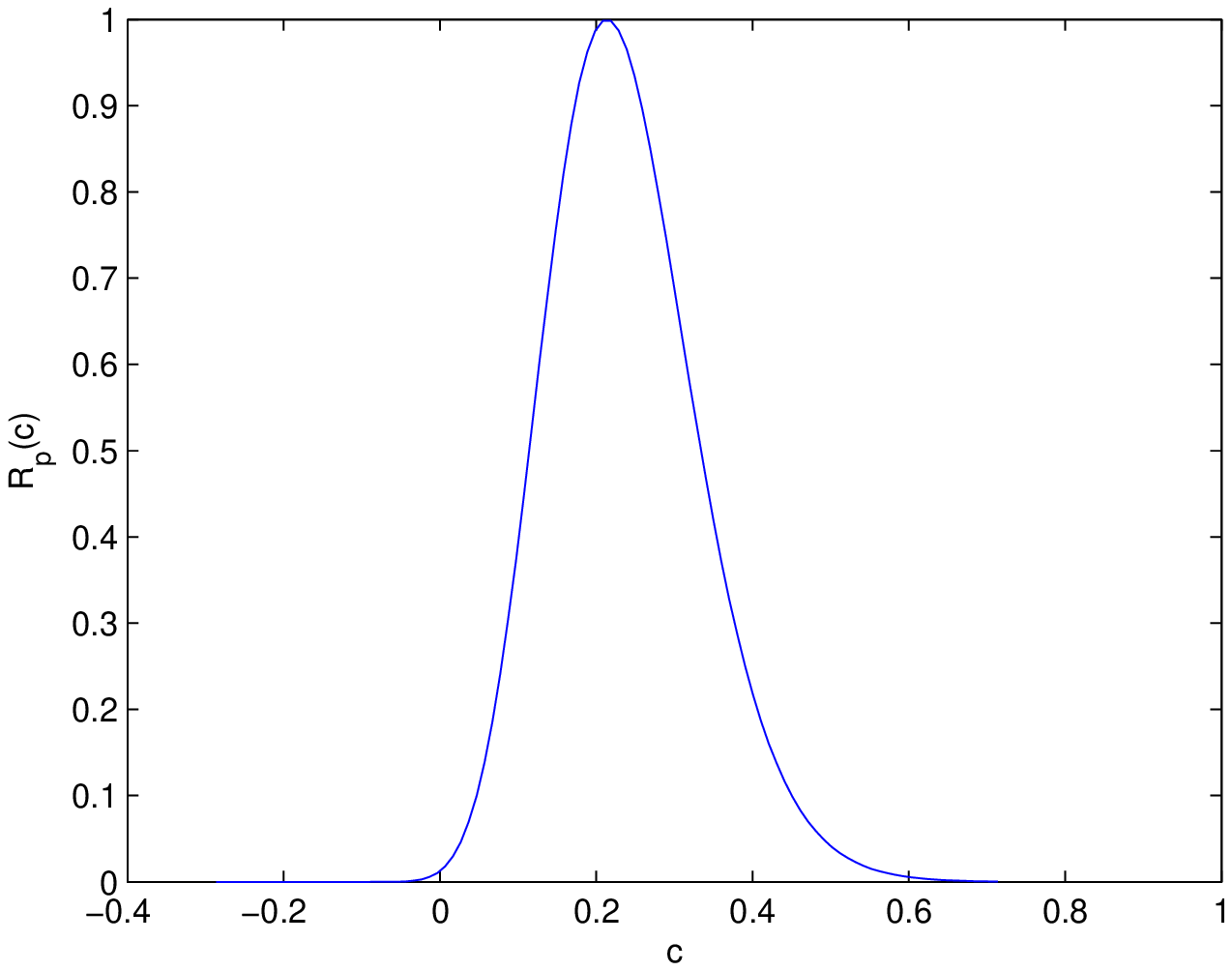}} \subfloat[]{%
\includegraphics[height=6cm]{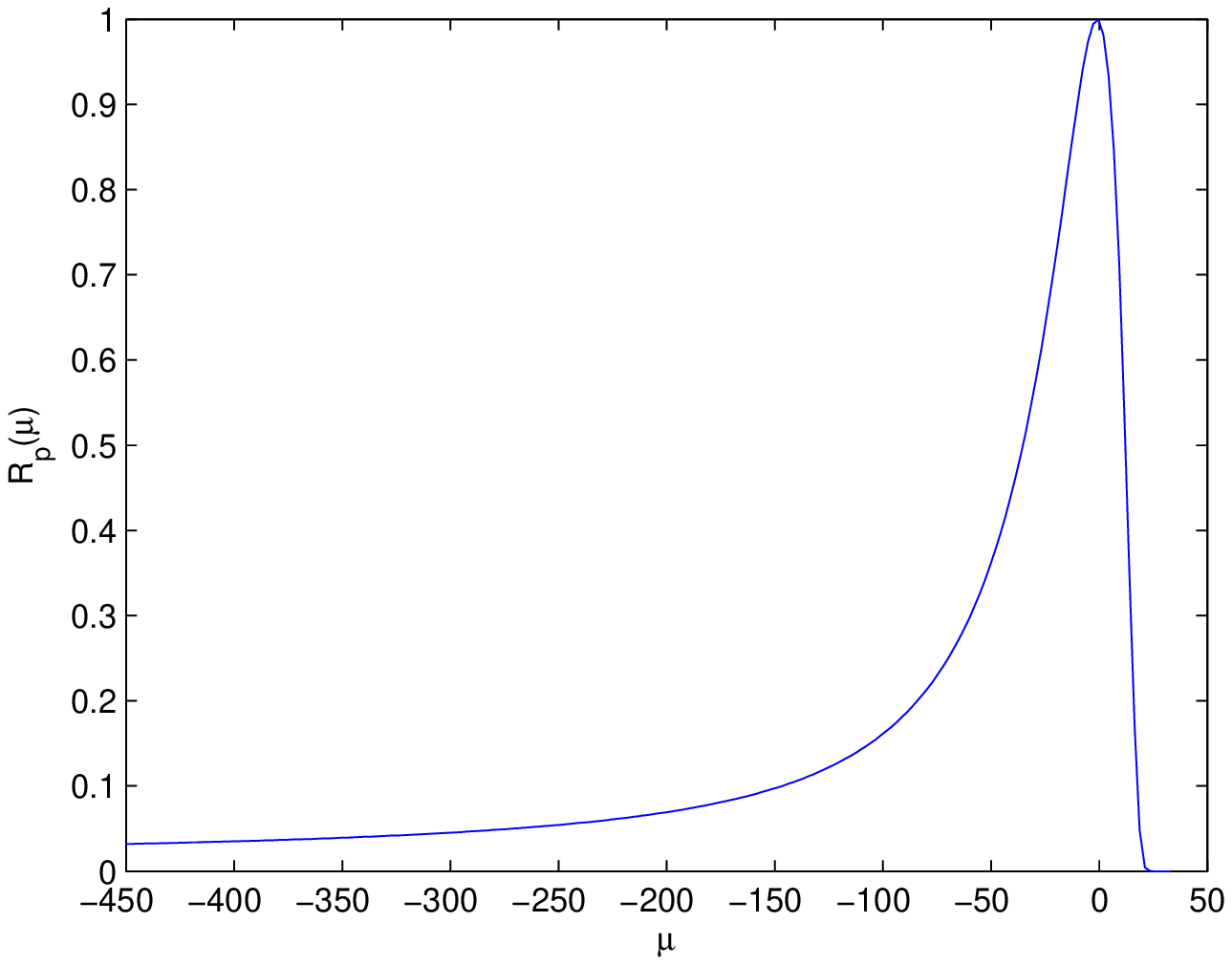}}
\end{center}
\caption{Rain data example: (a) Relative profile likelihood of GEV
shape parameter $c.$ (b)
Relative profile likelihood of threshold parameter in three parameter Fr\'{e}%
chet model.} \label{fig7}
\end{figure}

In the state of Michoac\'{a}n, M\'{e}xico, near its capital city
Morelia, there is a monitoring meteorological station located at the
Cointzio dam. This station is representative of rainfall patterns in
this area. Yearly maxima of daily rainfall were obtained for 58
years in a period between 1940 and 2002. In this area, there is a
marked rainy season from May to September. This data set will serve
to illustrate the statistical modelling procedures suggested here.
As a first step, the relative profile likelihood of the GEV shape
parameter $c$ shown in Figure \ref{fig7}(a) assigns plausibility
only to positive values of $c$ and the mle is $\hat{c}=0.21.$
therefore suggesting a Fr\'{e}chet model. Since rain data are
necessarily non-negative, for physical reasons it is important to
consider a Fr\'{e}chet model with a non-negative lower threshold
parameter $\mu \geq 0$ that could very well simplify to a two
parameter Fr\'{e}chet model, where $\mu =0$. The relative profile
likelihood of $\mu $ under the three parameter Fr\'{e}chet model
shown in Figure \ref{fig7}(b), clearly assigns a very high
plausibility to the value of $\mu =0,$ so that the data appear to
support strongly a two parameter Fr\'{e}chet model. Under this
model, the maximum likelihood
estimates are%
\begin{equation*}
\begin{tabular}{|c|c|c|c|}
\hline
$\hat{\sigma}$ & $\hat{\beta}$ & $\hat{Q}_{.95}$ & $\hat{Q}_{.99}$ \\ \hline
36.99 & 4.57 & 70.87 & 101.25 \\ \hline
\end{tabular}%
.
\end{equation*}

\begin{figure}[tbp]
\begin{center}
\subfloat[]{\includegraphics[height=6cm]{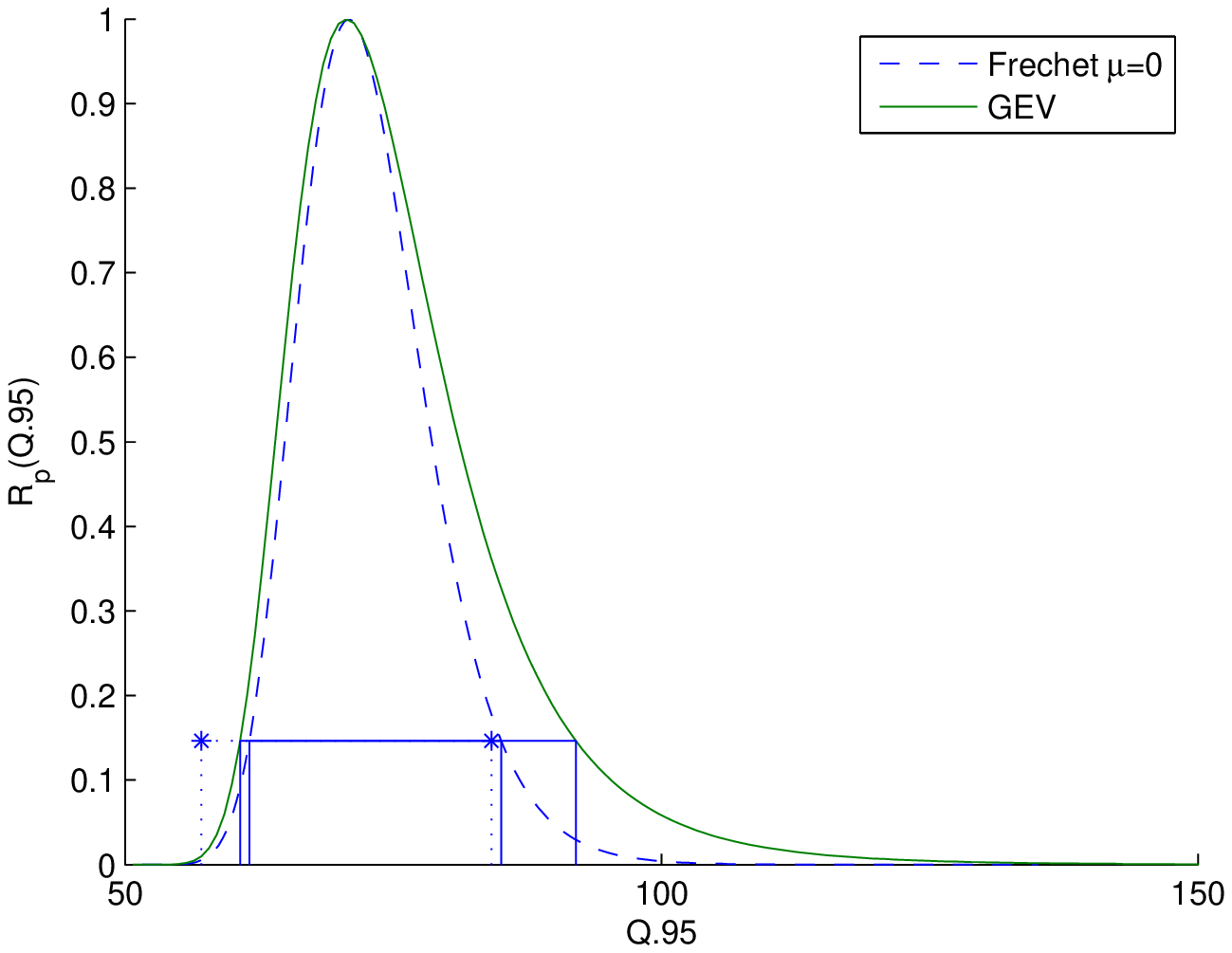}} \subfloat[]{%
\includegraphics[height=6cm]{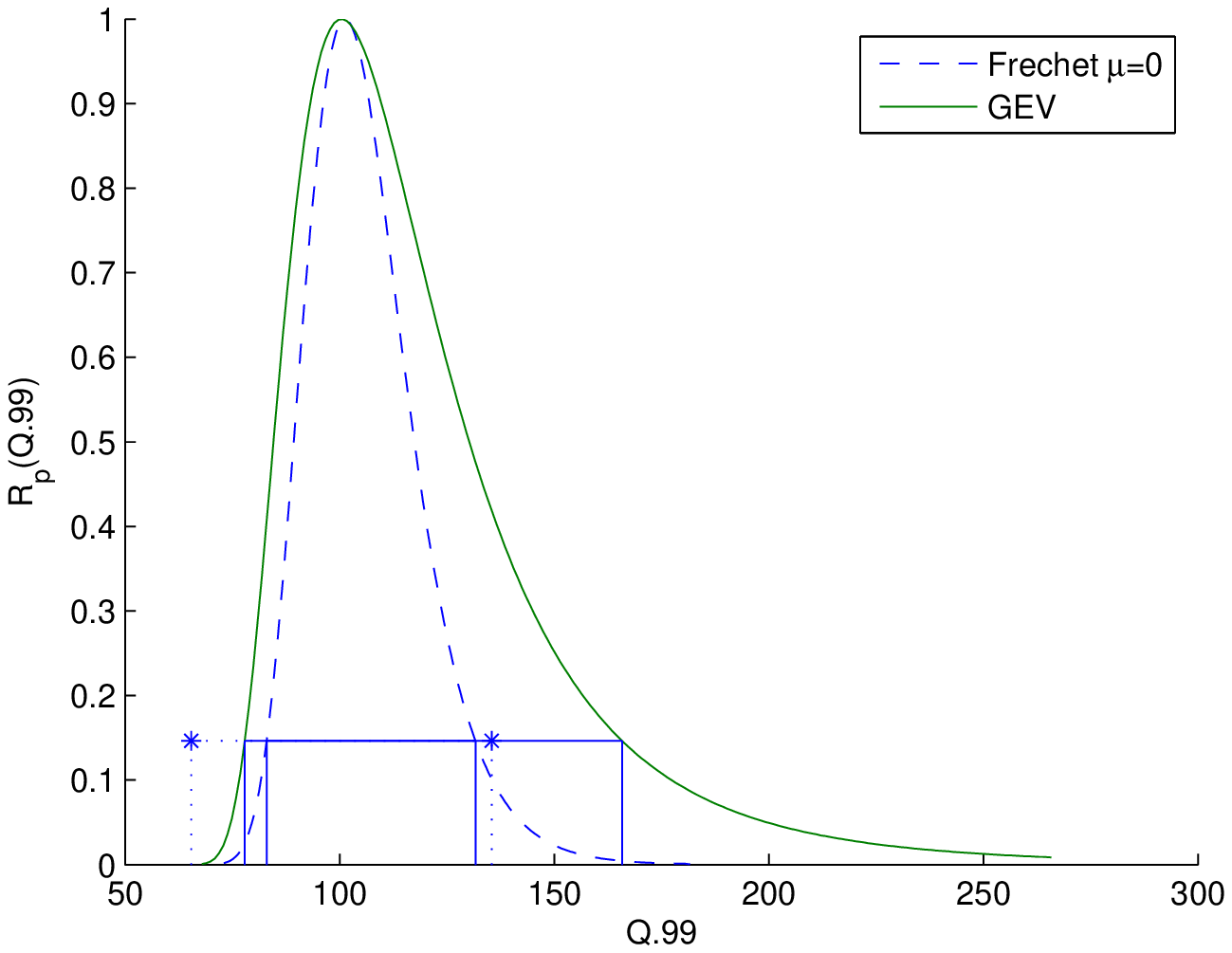}}
\end{center}
\caption{Rain data example:  Relative profile likelihood of (a)
$Q_{.95}$, (b) $Q_{.99}.$ } \label{fig8}
\end{figure}

Figures \ref{fig8}(a) and \ref{fig8}(b) present together, for the
sake of comparison, the corresponding relative profile likelihoods
of these large quantiles of interest under the two parameter
Fr\'{e}chet model and also under the GEV model without any
restrictions to its parameters. The GEV model without
restrictions for its threshold corresponds as well to a three parameter Fr%
\'{e}chet model without restriction to its threshold parameter; the
corresponding Fr\'{e}chet mle's are%
\begin{equation*}
\begin{tabular}{|c|c|c|c|c|}
\hline
$\hat{\mu}$ & $\hat{\sigma}$ & $\hat{\beta}$ & $\hat{Q}_{.95}$ & $\hat{Q}%
_{.99}$ \\ \hline
-1.55 & 38.57 & 4.76 & 70.64 & 100.44 \\ \hline
\end{tabular}%
.
\end{equation*}%
In terms of the GEV distribution's parameters, the mle's are given by

\begin{equation*}
\begin{tabular}{|c|c|c|}
\hline
$\hat{a}$ & $\hat{b}$ & $\hat{c}$ \\ \hline
37.02 & 8.1 & 0.21 \\ \hline
\end{tabular}%
\end{equation*}

\bigskip The likelihood intervals obtained for these quantiles with the GEV
model are larger and imply that larger values of these quantiles are
plausible. Also in these graphs, the aml GEV intervals are marked and show
that their right endpoints tend to coincide with the right endpoints of the
profile likelihood intervals of the two Fr\'{e}chet model for these
quantiles; nevertheless the left points are much smaller than the other
likelihoods endpoints and therefore include small values of the quantiles
that are implausible under both models (two parameter Fr\'{e}chet and the
GEV). That is, the aml intervals tend to underestimate the values of the
quantiles.

\begin{figure}[tbp]
\begin{center}
\subfloat[]{\includegraphics[height=6cm]{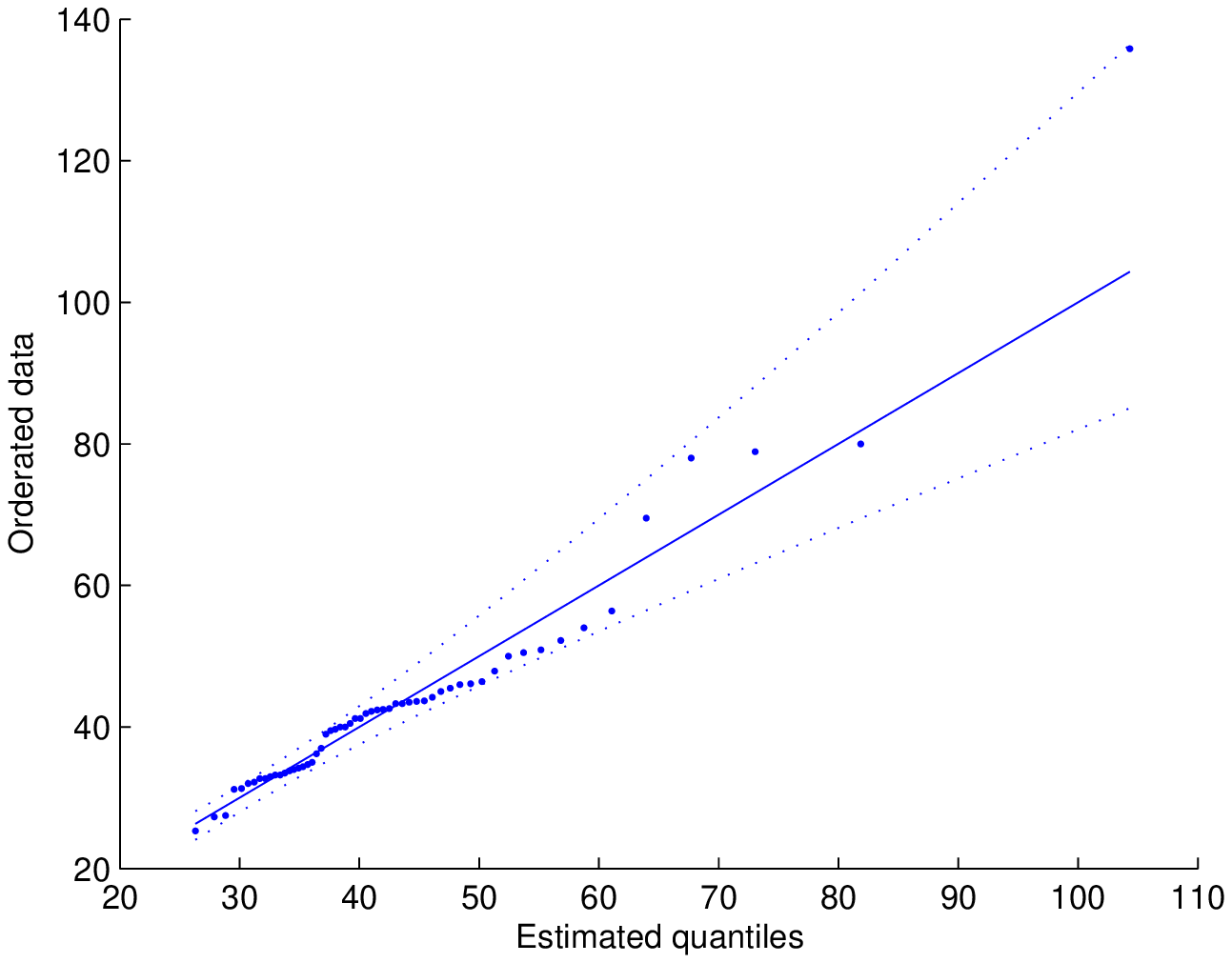}} \subfloat[]{%
\includegraphics[height=6cm]{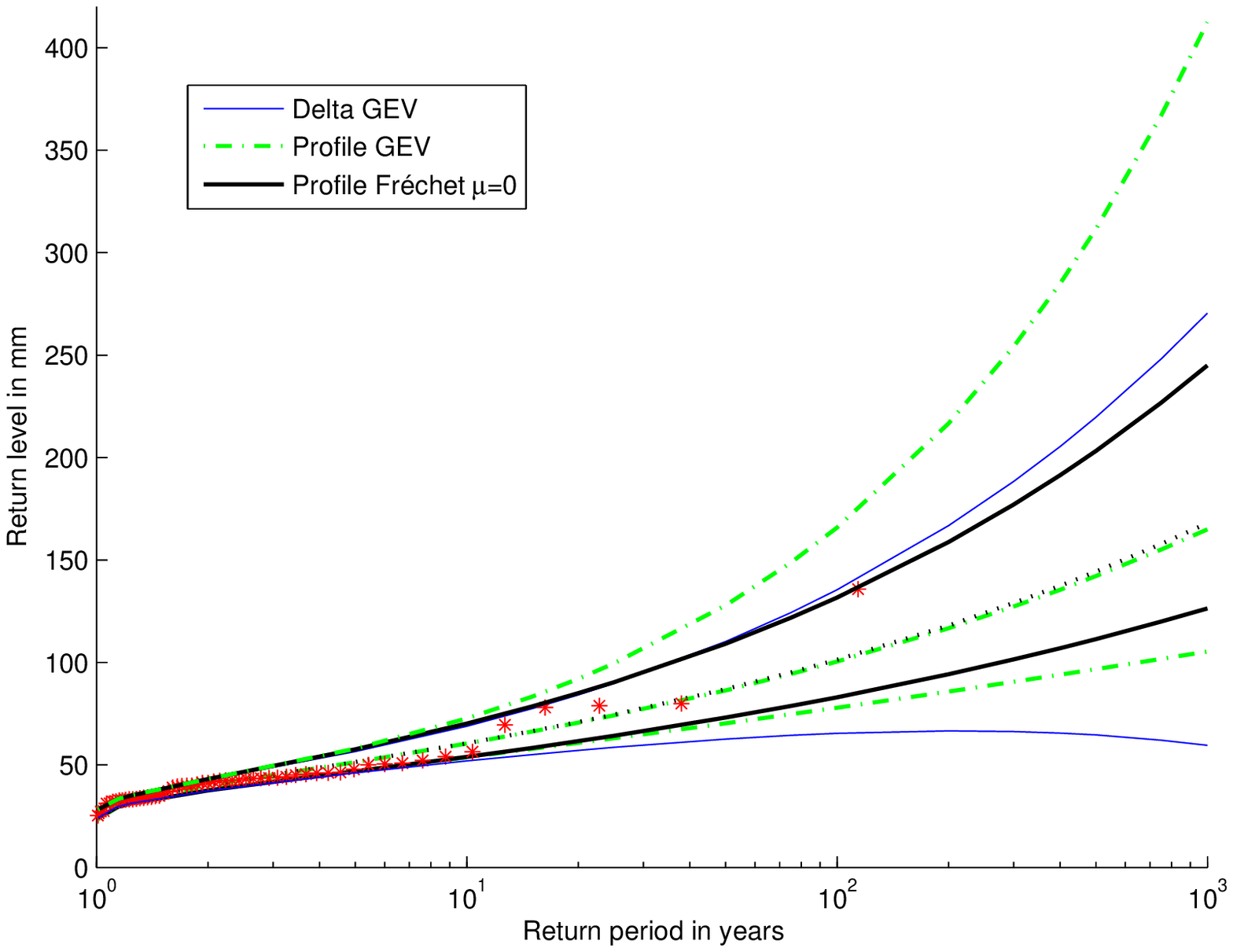}}
\end{center}
\caption{Rain data example: (a) Q-Q plot for the two parameter
Fr\'{e}chet model. (b) Return period plot.} \label{fig9}
\end{figure}

The likelihood ratio statistic of these two models for this data set is%
\begin{equation*}
W = \frac{L_{\text{Fr\'{e}chet}}\left( \mu =0,\hat{\sigma},\hat{\beta};x\right)
}{L_{\text{Fr\'{e}chet}}\left( \hat{\mu},\hat{\sigma},\hat{\beta};x\right) }%
=0.9983.
\end{equation*}%
Since these models are nested, the observed value of $-2\log W = 0.0034$ has
$p$-value of $0.9535$ under the asymptotic  chi-square distribution with one degree
of freedom. The observed value of $0.9983$ with a p-value of 0.32, indicates that the
two Fr\'{e}chet parameter model makes the observed sample equally probable. However
since the two Fr\'{e}chet parameter model is simpler and fits adequately the data set
as shown in Figure \ref{fig9}(a), this model should be preferred. Figure
\ref{fig9}(a) shows the corresponding quantile-quantile plot with pointwise
likelihood bands that includes all observed values. Moreover, this model should be
taken into account due to the physical considerations stated above.

Likelihood-confidence intervals of 15\% likelihood level and
approximate 95\% confidence level for the quantiles of interest
$Q_{.95}$ and $Q_{.99}$ under the two parameter Fr\'{e}chet model
are $\left( 61.6,85.06\right) $ and $\left( 83.02,131.66\right)$
respectively. Finally Figure \ref{fig9}(b) shows the return periods
plot with profile likelihood 15\% level bands marked for both the
GEV model and the two Fr\'{e}chet model. Since rainfall levels
higher than 200ml are associated with floodings of Morelia, and
since a return period of a 100 years is associated to quantile
$Q_{.99},$ then the probability is extremely low that the city of
Morelia gets flooded within 100 years.

\section{Conclusions}

Overall, profile likelihood intervals of large quantiles of Extreme Value
distributions and of the GEV shape parameter $c$ performed well and had
adequate coverage frequencies for moderate and small sample sizes. In
contrast, the corresponding aml intervals are symmetric about the mle and
had lower and poor coverage frequencies in the case of samples of size $%
n\leq 100.$ Moreover, a large proportion of the aml intervals that excluded
the true value tended to underestimate it. The aml intervals are frequently
used in Extreme Value Theory applications without notice of these issues.

Profile likelihood intervals of EV submodels tend to be shorter than the
corresponding GEV profile likelihood intervals when the true value of $c$ is
close to zero, that is when $c\in \left( -.05,.05\right) $ if the sample
size is $n\leq 50$. Nevertheless, their coverage frequencies are adequate so
that they should be preferred when the model selection of an EV is clear.
However, if there is no additional external information on a given preferred
EV model suggested by the theory behind the specific phenomenon of interest,
then using GEV profile likelihood intervals is a conservative procedure
since they also had good coverage frequencies, even though these intervals
tended to be larger.

Profile likelihood intervals of $c$ may serve as an aid in model selection.
They also had adequate coverage frequencies. For values of $c$ in a region
around zero $\left( -0.01,0.01\right) $ approximately 95\% of the likelihood
intervals for the simulated samples included the value of zero. \ These are
cases where the three EV models are plausible for the given sample, and also
where the Gumbel model usually has a moderate or high plausibility given by
the relative profile likelihood of $c$ at zero. This is indicative of the
need of additional external information of experts and other diagnostic
methods to select adequately the best and most simple model for the
phenomenon of interest. This will improve the estimating precision, and will
prevent underestimating the quantile of interest.

Finally, for sample sizes smaller than 50 and in the case that a Weibull
model might be an appropriate choice$,$ then the use of the exact likelihood
function is suggested in order to make inferences about the parameters of
interest through profile likelihood intervals.

\section{Acknowledgments}

Work partially financed by CONCYTEG, Grants 05-02-K117-027-A04 and
05-02-K117-099. The authors thank Instituto Mexicano de Tecnolog\'{\i}a del
Agua at Cuernavaca, M\'{e}xico, for facilitating the rainfall data set.

\begin{center}
{\scriptsize 
{%
\begin{tabular}{|c||c|c|c|c|c|c||c|c|c|c|c|c|c|}
\hline
\multicolumn{14}{|c|}{n=100, $Q_{95}$} \\ \hline
& \multicolumn{6}{|c||}{SUBMODEL} & \multicolumn{7}{|c|}{GEV} \\ \hline
& \multicolumn{3}{|c|}{Profile Likelihood Ints.} & \multicolumn{3}{|c||}{AML}
& \multicolumn{3}{|c|}{Profile Likelihood Ints.} &  & \multicolumn{3}{|c|}{
AML} \\ \hline
c & $<$~ & C. F. & $>$~ & $<$~ & C. F. & $>$~ & $<$~ & C. F. & $>$~ & SNP & $%
<$~ & C. F. & $>$~ \\ \hline
-0.5 & 543 & 9314 & 143 & 984 & 8877 & 139 & 543 & 9314 & 143 & 0 & 984 &
8877 & 139 \\ \hline
-0.4 & 438 & 9408 & 154 & 884 & 9003 & 113 & 438 & 9408 & 154 & 0 & 884 &
9003 & 113 \\ \hline
-0.3 & 442 & 9374 & 184 & 888 & 9023 & 89 & 442 & 9374 & 184 & 0 & 887 & 9024
& 89 \\ \hline
-0.2 & 379 & 9449 & 172 & 872 & 9076 & 52 & 378 & 9450 & 172 & 0 & 872 & 9076
& 52 \\ \hline
-0.1 & 347 & 9457 & 196 & 844 & 9120 & 36 & 344 & 9469 & 187 & 0 & 842 & 9124
& 34 \\ \hline
-0.05 & 392 & 9423 & 185 & 805 & 9169 & 26 & 363 & 9464 & 173 & 0 & 805 &
9170 & 25 \\ \hline
-0.01 & 400 & 9395 & 205 & 755 & 9215 & 30 & 315 & 9496 & 189 & 0 & 755 &
9216 & 29 \\ \hline
-0.001 & 423 & 9374 & 203 & 778 & 9191 & 31 & 338 & 9469 & 193 & 0 & 777 &
9192 & 31 \\ \hline
0.0 & 394 & 9415 & 191 & 751 & 9221 & 28 & 325 & 9490 & 185 & 0 & 751 & 9221
& 28 \\ \hline
0.001 & 416 & 9380 & 204 & 792 & 9180 & 28 & 332 & 9473 & 195 & 0 & 792 &
9182 & 26 \\ \hline
0.01 & 443 & 9396 & 161 & 785 & 9194 & 21 & 341 & 9509 & 150 & 0 & 784 & 9196
& 20 \\ \hline
0.05 & 446 & 9358 & 196 & 735 & 9240 & 25 & 309 & 9501 & 190 & 0 & 731 & 9245
& 24 \\ \hline
0.1 & 496 & 9302 & 202 & 781 & 9195 & 24 & 321 & 9477 & 202 & 0 & 781 & 9195
& 24 \\ \hline
0.2 & 285 & 9495 & 220 & 727 & 9262 & 11 & 267 & 9513 & 220 & 0 & 727 & 9262
& 11 \\ \hline
0.3 & 280 & 9506 & 214 & 757 & 9232 & 11 & 280 & 9506 & 214 & 0 & 757 & 9232
& 11 \\ \hline
0.4 & 298 & 9477 & 225 & 782 & 9214 & 04 & 298 & 9477 & 225 & 0 & 782 & 9214
& 04 \\ \hline
0.5 & 296 & 9472 & 232 & 740 & 9260 & 0 & 296 & 9472 & 232 & 0 & 740 & 9260
& 0 \\ \hline\hline
\multicolumn{14}{|c|}{n=50} \\ \hline
c & $<$~ & C. F. & $>$~ & $<$~ & C. F. & $>$~ & $<$~ & C. F. & $>$~ & SNP & $%
<$~ & C. F. & $>$~ \\ \hline
-0.5 & 576 & 9288 & 136 & 1382 & 845 & 168 & 572 & 9287 & 136 & 5 & 1379 &
8448 & 168 \\ \hline
-0.4 & 552 & 9309 & 139 & 1346 & 8547 & 107 & 551 & 9308 & 139 & 2 & 1343 &
8548 & 107 \\ \hline
-0.3 & 505 & 9345 & 15 & 1271 & 8663 & 66 & 505 & 9345 & 15 & 0 & 1271 & 8663
& 66 \\ \hline
-0.2 & 433 & 9435 & 132 & 1195 & 8763 & 42 & 433 & 9435 & 132 & 0 & 1195 &
8763 & 42 \\ \hline
-0.1 & 412 & 9432 & 156 & 1105 & 8881 & 14 & 383 & 9466 & 151 & 0 & 1105 &
8882 & 13 \\ \hline
-0.05 & 447 & 9379 & 174 & 1111 & 8875 & 14 & 388 & 9449 & 163 & 0 & 1112 &
8876 & 12 \\ \hline
0.0 & 503 & 9327 & 17 & 1047 & 8945 & 8 & 351 & 9486 & 163 & 0 & 1045 & 8947
& 8 \\ \hline
0.05 & 567 & 9243 & 19 & 1067 & 8927 & 6 & 346 & 9466 & 188 & 0 & 1063 & 8931
& 6 \\ \hline
0.1 & 64 & 9168 & 192 & 1069 & 8929 & 2 & 359 & 9449 & 192 & 0 & 1068 & 893
& 2 \\ \hline
0.2 & 461 & 9333 & 206 & 1001 & 8997 & 2 & 304 & 9491 & 205 & 0 & 1001 & 8997
& 2 \\ \hline
0.3 & 333 & 9458 & 209 & 978 & 9022 & 0 & 297 & 9494 & 209 & 0 & 977 & 9023
& 0 \\ \hline
0.4 & 309 & 9492 & 199 & 985 & 9015 & 0 & 307 & 9494 & 199 & 0 & 985 & 9015
& 0 \\ \hline
0.5 & 283 & 9468 & 249 & 993 & 9007 & 0 & 283 & 9468 & 249 & 0 & 993 & 9007
& 0 \\ \hline\hline
\multicolumn{14}{|c|}{n=25} \\ \hline
c & $<$~ & C. F. & $>$~ & $<$~ & C. F. & $>$~ & $<$~ & C. F. & $>$~ & SNP & $%
<$~ & C. F. & $>$~ \\ \hline
-0.5 & 757 & 9107 & 136 & 2165 & 755 & 285 & 583 & 8787 & 115 & 515 & 1826 &
7469 & 19 \\ \hline
-0.4 & 639 & 9244 & 117 & 2037 & 7834 & 129 & 55 & 9153 & 109 & 188 & 1886 &
7815 & 111 \\ \hline
-0.3 & 599 & 9287 & 114 & 1839 & 8097 & 64 & 545 & 9264 & 112 & 79 & 1773 &
8089 & 59 \\ \hline
-0.2 & 556 & 9313 & 131 & 1759 & 8209 & 32 & 526 & 9328 & 129 & 17 & 1744 &
8208 & 31 \\ \hline
-0.1 & 547 & 9315 & 138 & 1658 & 833 & 12 & 499 & 9356 & 136 & 9 & 1648 &
8331 & 12 \\ \hline
-0.05 & 595 & 9259 & 146 & 1659 & 8334 & 7 & 485 & 9365 & 145 & 5 & 1656 &
8332 & 7 \\ \hline
0 & 586 & 9252 & 162 & 1535 & 8457 & 8 & 398 & 9439 & 158 & 5 & 1528 & 8459
& 8 \\ \hline
0.05 & 714 & 9103 & 183 & 1576 & 8421 & 3 & 43 & 9385 & 182 & 3 & 1571 & 8423
& 3 \\ \hline
0.1 & 734 & 9109 & 157 & 1481 & 8518 & 1 & 382 & 9458 & 157 & 3 & 1478 & 8518
& 1 \\ \hline
0.2 & 819 & 9008 & 173 & 1457 & 8543 & 0 & 392 & 9435 & 173 & 0 & 1457 & 8543
& 0 \\ \hline
0.3 & 625 & 9195 & 18 & 1434 & 8566 & 0 & 338 & 9482 & 18 & 0 & 1433 & 8567
& 0 \\ \hline
0.4 & 398 & 9414 & 188 & 1311 & 8689 & 0 & 286 & 9526 & 188 & 0 & 1311 & 8689
& 0 \\ \hline
0.5 & 353 & 9434 & 213 & 1385 & 8615 & 0 & 307 & 948 & 213 & 0 & 1385 & 8615
& 0 \\ \hline
\end{tabular}%
}} \label{tab3} 

{\small Table 3. Coverage frequencies for $Q_{.95}$ with sample sizes 100,
50 and 25. C.F. stands for Coverage Frequencies, `$<$' is the number of
intervals that fell below the true value, `$>$' the number that fell above
and SNP represents the number of samples with numerical problems.}
\end{center}

\bigskip

\begin{center}
{\scriptsize 
{%
\begin{tabular}{|c||c|c|c|c|c|c||c|c|c|c|c|c|c|}
\hline
\multicolumn{14}{|c|}{n=100, $Q_{99}$} \\ \hline
& \multicolumn{6}{|c||}{SUBMODEL} & \multicolumn{7}{|c|}{GEV} \\ \hline
& \multicolumn{3}{|c|}{Profile Likelihood Ints.} & \multicolumn{3}{|c||}{AML}
& \multicolumn{3}{|c|}{Profile Likelihood Ints.} &  & \multicolumn{3}{|c|}{
AML} \\ \hline
c & $<$~ & C. F. & $>$~ & $<$~ & C. F. & $>$~ & $<$~ & C. F. & $>$~ & SNP & $%
<$~ & C. F. & $>$~ \\ \hline
-0.5 & 590 & 9332 & 78 & 1730 & 8258 & 12 & 590 & 9332 & 78 & 0 & 1731 & 8257
& 12 \\ \hline
-0.4 & 448 & 9460 & 92 & 1376 & 8622 & 2 & 448 & 9460 & 92 & 0 & 1375 & 8623
& 2 \\ \hline
-0.3 & 418 & 9440 & 142 & 1227 & 8770 & 3 & 418 & 9440 & 142 & 0 & 1227 &
8770 & 3 \\ \hline
-0.2 & 401 & 9444 & 155 & 1128 & 8870 & 2 & 401 & 9448 & 151 & 0 & 1128 &
8870 & 2 \\ \hline
-0.1 & 335 & 9376 & 289 & 1054 & 8940 & 6 & 335 & 9493 & 172 & 0 & 1053 &
8943 & 4 \\ \hline
-0.05 & 353 & 9325 & 322 & 967 & 9028 & 5 & 341 & 9479 & 180 & 0 & 966 & 9031
& 3 \\ \hline
-0.01 & 431 & 9286 & 283 & 913 & 9083 & 4 & 299 & 9490 & 211 & 0 & 913 & 9084
& 3 \\ \hline
-0.001 & 494 & 9242 & 264 & 923 & 9074 & 3 & 330 & 9468 & 202 & 0 & 923 &
9075 & 2 \\ \hline
0.0 & 490 & 9264 & 246 & 923 & 9077 & 0 & 324 & 9476 & 200 & 0 & 924 & 9076
& 0 \\ \hline
0.001 & 510 & 9247 & 243 & 930 & 9065 & 5 & 349 & 9458 & 193 & 0 & 930 & 9069
& 1 \\ \hline
0.01 & 570 & 9223 & 207 & 907 & 9091 & 2 & 337 & 9485 & 178 & 0 & 905 & 9094
& 1 \\ \hline
0.05 & 863 & 8913 & 224 & 875 & 9124 & 1 & 293 & 9490 & 217 & 0 & 869 & 9130
& 1 \\ \hline
0.1 & 857 & 8925 & 218 & 887 & 9113 & 0 & 315 & 9467 & 218 & 0 & 883 & 9117
& 0 \\ \hline
0.2 & 282 & 9486 & 232 & 845 & 9155 & 0 & 263 & 9505 & 232 & 0 & 845 & 9155
& 0 \\ \hline
0.3 & 269 & 9496 & 235 & 857 & 9143 & 0 & 269 & 9496 & 235 & 0 & 857 & 9143
& 0 \\ \hline
0.4 & 291 & 9472 & 237 & 887 & 9113 & 0 & 291 & 9472 & 237 & 0 & 888 & 9112
& 0 \\ \hline
0.5 & 288 & 9477 & 235 & 865 & 9135 & 0 & 288 & 9477 & 235 & 0 & 864 & 9136
& 0 \\ \hline\hline
\multicolumn{14}{|c|}{n=50} \\ \hline
& \multicolumn{6}{|c||}{SUBMODEL} & \multicolumn{7}{|c|}{GEV} \\ \hline
-0.5 & 628 & 9306 & 66 & 2297 & 7690 & 13 & 623 & 9306 & 66 & 5 & 2293 & 7689
& 13 \\ \hline
-0.4 & 576 & 9355 & 69 & 1915 & 8083 & 2 & 575 & 9354 & 69 & 2 & 1913 & 8083
& 2 \\ \hline
-0.3 & 521 & 9356 & 123 & 1702 & 8298 & 0 & 521 & 9359 & 120 & 0 & 1702 &
8298 & 0 \\ \hline
-0.2 & 429 & 9412 & 159 & 1475 & 8525 & 0 & 429 & 9452 & 119 & 0 & 1475 &
8525 & 0 \\ \hline
-0.1 & 555 & 9177 & 268 & 1399 & 8598 & 3 & 399 & 9440 & 161 & 0 & 1398 &
8602 & 0 \\ \hline
-0.05 & 421 & 9320 & 259 & 1326 & 8672 & 2 & 383 & 9460 & 157 & 0 & 1325 &
8675 & 0 \\ \hline
0.0 & 624 & 9157 & 219 & 1249 & 8750 & 1 & 361 & 9459 & 180 & 0 & 1249 & 8751
& 0 \\ \hline
0.05 & 923 & 8872 & 205 & 1261 & 8739 & 0 & 348 & 9455 & 197 & 0 & 1257 &
8743 & 0 \\ \hline
0.1 & 1131 & 8661 & 208 & 1232 & 8768 & 0 & 333 & 9460 & 207 & 0 & 1231 &
8769 & 0 \\ \hline
0.2 & 577 & 9190 & 233 & 1138 & 8862 & 0 & 289 & 9478 & 233 & 0 & 1137 & 8863
& 0 \\ \hline
0.3 & 313 & 9459 & 228 & 1127 & 8873 & 0 & 281 & 9491 & 228 & 0 & 1127 & 8873
& 0 \\ \hline
0.4 & 286 & 9479 & 235 & 1172 & 8828 & 0 & 286 & 9479 & 235 & 0 & 1173 & 8827
& 0 \\ \hline
0.5 & 264 & 9461 & 275 & 1137 & 8863 & 0 & 264 & 9461 & 275 & 0 & 1136 & 8864
& 0 \\ \hline\hline
\multicolumn{14}{|c|}{n=25} \\ \hline
& \multicolumn{6}{|c||}{SUBMODEL} & \multicolumn{7}{|c|}{GEV} \\ \hline
-0.5 & 772 & 9166 & 62 & 3069 & 6896 & 35 & 530 & 8893 & 62 & 515 & 2600 &
6872 & 13 \\ \hline
-0.4 & 668 & 9247 & 85 & 2687 & 7309 & 4 & 549 & 9199 & 64 & 188 & 2509 &
7299 & 4 \\ \hline
-0.3 & 624 & 9221 & 155 & 2321 & 7678 & 1 & 564 & 9257 & 100 & 79 & 2244 &
7677 & 0 \\ \hline
-0.2 & 531 & 9290 & 179 & 2078 & 7922 & 0 & 516 & 9354 & 113 & 17 & 2061 &
7922 & 0 \\ \hline
-0.1 & 535 & 9250 & 215 & 1916 & 8083 & 1 & 497 & 9349 & 145 & 9 & 1907 &
8084 & 0 \\ \hline
-0.05 & 572 & 9234 & 194 & 1901 & 8099 & 0 & 458 & 9389 & 148 & 5 & 1896 &
8099 & 0 \\ \hline
0.0 & 659 & 9149 & 192 & 1755 & 8245 & 0 & 405 & 9437 & 153 & 5 & 1748 & 8247
& 0 \\ \hline
0.05 & 1015 & 8790 & 195 & 1822 & 8178 & 0 & 418 & 9396 & 183 & 3 & 1817 &
8180 & 0 \\ \hline
0.1 & 1252 & 8555 & 193 & 1639 & 8361 & 0 & 377 & 9431 & 189 & 3 & 1636 &
8361 & 0 \\ \hline
0.2 & 1363 & 8435 & 202 & 1618 & 8382 & 0 & 389 & 9409 & 202 & 0 & 1616 &
8384 & 0 \\ \hline
0.3 & 788 & 8999 & 213 & 1607 & 8393 & 0 & 330 & 9457 & 213 & 0 & 1606 & 8394
& 0 \\ \hline
0.4 & 411 & 9357 & 232 & 1492 & 8508 & 0 & 286 & 9482 & 232 & 0 & 1492 & 8508
& 0 \\ \hline
0.5 & 326 & 9414 & 260 & 1593 & 8407 & 0 & 298 & 9442 & 260 & 0 & 1593 & 8407
& 0 \\ \hline
\end{tabular}%
}}     \label{tab4}

{\small Table 4. Coverage frequencies for $Q_{.99}$ with sample sizes 100,
50 and 25. C.F. stands for Coverage Frequencies, `$<$' is the number of
intervals that fell below the true value, `$>$' the number that fell above
and SNP represents the number of samples with numerical problems}.

{\footnotesize {%
\begin{tabular}{|c|c|c|c|c|c|}
\hline
\multicolumn{6}{|c|}{15\% Profile Likelihood Intervals for c with n=100} \\
\hline
c & $<$ & Cov. Freq. & $>$ & Correct & Negative \\ \hline
-0.5 & 564 & 9328 & 108 & 10000 & 0 \\ \hline
-0.4 & 396 & 9479 & 125 & 10000 & 0 \\ \hline
-0.3 & 410 & 9425 & 165 & 10000 & 45 \\ \hline
-0.2 & 371 & 9451 & 178 & 9993 & 1721 \\ \hline
-0.1 & 348 & 9465 & 187 & 9375 & 6988 \\ \hline
-0.05 & 297 & 9458 & 245 & 7787 & 8766 \\ \hline
-0.01 & 272 & 9490 & 238 & 5772 & 9416 \\ \hline
-0.001 & 296 & 9465 & 239 & 5264 & 9458 \\ \hline
0.0 & 303 & 9460 & 237 & 0 & 9460 \\ \hline
0.001 & 309 & 9461 & 230 & 4735 & 9454 \\ \hline
0.01 & 286 & 9483 & 231 & 5357 & 9392 \\ \hline
0.05 & 255 & 9482 & 263 & 7299 & 8605 \\ \hline
0.1 & 299 & 9444 & 257 & 8866 & 6773 \\ \hline
0.2 & 244 & 9463 & 293 & 9902 & 2232 \\ \hline
0.3 & 246 & 9465 & 289 & 9992 & 265 \\ \hline
0.4 & 236 & 9483 & 281 & 10000 & 5 \\ \hline
0.5 & 259 & 9467 & 274 & 10000 & 0 \\ \hline\hline
\multicolumn{6}{|c|}{15\% Profile Likelihood Intervals for c with n=50} \\
\hline
c & $<$ & Cov. Freq. & $>$ & Correct & Negative \\ \hline
-0.3 & 467 & 9394 & 139 & 9989 & 1653 \\ \hline
-0.2 & 388 & 9441 & 171 & 9821 & 5015 \\ \hline
-0.1 & 371 & 9416 & 213 & 8515 & 8206 \\ \hline
-0.05 & 327 & 9466 & 207 & 7075 & 8996 \\ \hline
0.0 & 317 & 9442 & 241 & 0 & 9442 \\ \hline
0.05 & 287 & 9456 & 257 & 6445 & 8987 \\ \hline
0.1 & 321 & 9419 & 260 & 7939 & 7917 \\ \hline
0.2 & 255 & 9460 & 285 & 9426 & 5022 \\ \hline
0.3 & 256 & 9463 & 281 & 9833 & 2276 \\ \hline
0.4 & 271 & 9458 & 271 & 9969 & 767 \\ \hline
0.5 & 246 & 9434 & 320 & 9988 & 158 \\ \hline
\end{tabular}%
} \label{tab5} }

\medskip {\small Table 5. Coverage frequencies for $c$ with sample sizes 100
and 50: \lq$<$' is the number of intervals that fell below the true value,
\lq$>$' the number that fell above, \lq Correct' stands for the number of
samples with correct choice of EV and \lq Negative' stands for the number of
samples with negative product of interval endpoints.}
\end{center}

\end{document}